\documentclass[aps,prx,twocolumn,superscriptaddress,longbibliography]{revtex4-1}

\usepackage{lipsum}
\usepackage[dvipsnames]{xcolor}
\usepackage{bm}
\usepackage{graphicx}
\usepackage{upgreek}
\usepackage{natbib}
\usepackage{braket}
\usepackage{physics}

\usepackage[%
breaklinks,%
pdftex,%
hyperfootnotes=true,%
pdfpagelabels,%
bookmarks,%
pageanchor,%
]{hyperref}

\hypersetup{%
	colorlinks=true, linktocpage=true, pdfstartpage=1, pdfstartview=FitH, pdfborder={0 0 0},%
	breaklinks=true, pdfpagemode=UseNone, pageanchor=true, pdfpagemode=UseOutlines,%
	plainpages=false, bookmarksnumbered, bookmarksopen=true, bookmarksopenlevel=1,%
	hypertexnames=true, pdfhighlight=/O,
	urlcolor=Red!50!Sepia, linkcolor=NavyBlue, citecolor=RoyalBlue, 
}

\def\>{\right\rangle}
\def\<{\left\langle}
\def\be{\begin{equation}}
\def\ee{\end{equation}}
\def\ba{\begin{array}{lll}}
\def\ea{\end{array}}

\def\beq{\begin{eqnarray}}
\def\eeq{\end{eqnarray}}

\begin{document}

\title{Engineering dynamical couplings for quantum thermodynamic tasks}
\author{Matteo Carrega}
\affiliation{CNR-SPIN,  Via  Dodecaneso  33,  16146  Genova, Italy}
\email{matteo.carrega@spin.cnr.it}
    \author{Loris Maria Cangemi}
    \affiliation{Dipartimento di Fisica ``E.~Pancini'', Universit\`a di Napoli ``Federico II'', Complesso di Monte S.~Angelo, via Cinthia, 80126 Napoli, Italy}
    \affiliation{CNR-SPIN, c/o Complesso di Monte S. Angelo, via Cinthia - 80126 - Napoli, Italy}
    
    \author{Giulio De Filippis}
    \affiliation{Dipartimento di Fisica ``E.~Pancini'', Universit\`a di Napoli ``Federico II'', Complesso di Monte S.~Angelo, via Cinthia, 80126 Napoli, Italy}
    \affiliation{CNR-SPIN, c/o Complesso di Monte S. Angelo, via Cinthia - 80126 - Napoli, Italy}
    \affiliation{INFN, Sezione di Napoli, Complesso Universitario di Monte S. Angelo,
I-80126 Napoli, Italy}
\author{Vittorio Cataudella}
    \affiliation{Dipartimento di Fisica ``E.~Pancini'', Universit\`a di Napoli ``Federico II'', Complesso di Monte S.~Angelo, via Cinthia, 80126 Napoli, Italy}
    \affiliation{CNR-SPIN, c/o Complesso di Monte S. Angelo, via Cinthia - 80126 - Napoli, Italy}
    \affiliation{INFN, Sezione di Napoli, Complesso Universitario di Monte S. Angelo,
I-80126 Napoli, Italy}
\author{Giuliano Benenti}
    \affiliation{Center for Nonlinear and Complex Systems, Dipartimento di Scienza e Alta Tecnologia, Universit\`a degli Studi dell'Insubria, via Valleggio 11, 22100 Como, Italy} 
    \affiliation{Istituto Nazionale di Fisica Nucleare, Sezione di Milano, via Celoria 16, 20133 Milano, Italy}
    \affiliation{NEST, Istituto Nanoscienze-CNR, I-56126 Pisa, Italy}
    \author{Maura Sassetti}
    \affiliation{Dipartimento di Fisica, Universit\`a di Genova, Via Dodecaneso 33, 16146 Genova, Italy} 
    \affiliation{CNR-SPIN,  Via  Dodecaneso  33,  16146  Genova, Italy}

\begin{abstract}
{Describing the thermodynamic properties of quantum systems far from equilibrium is challenging, 
in particular when the system is strongly coupled to its environment, or when memory effects cannot be neglected.
Here, we address such regimes when the system-baths couplings are periodically modulated in time.
We show that the couplings modulation, usually associated to a purely dissipative effect when done 
nonadiabatically, can be suitably engineered
to perform thermodynamic tasks.
In particular, asymmetric couplings to two heat baths can be used to extract heat from the cold reservoir
and to realize an ideal heat rectifier, where the heat current can be blocked either in the forward or in the reverse 
configuration by simply tuning the frequency of the couplings modulation. 
Interestingly, both effects take place in the low-temperature, quantum non Markovian regime. 
Our work paves the way for the use of optimal control techniques for heat engines 
and refrigerators working in regimes beyond standard approaches.}
\end{abstract}

\maketitle

\section{Introduction}
\label{intro}
The development of quantum technologies~\cite{qcbook} requires a deeper understanding
of the thermodynamics of far from equilibrium quantum systems~\cite{esposito09,campisi11,kosloff13,Gelbwaser2015,Vinjanampathy2015,Sothmann2015,Goold2016,benenti17,talkner20,landi20,landi21,ciccarello21}.
Questions like how to efficiently manage heat at the nanoscale~\cite{Baowen2012,Benenti2016,Fornieri17,Pekola21}, what 
are the ultimate bounds to the performance of 
heat engines~\cite{benenti11,allahverdyan13,whitney14, ludovico14, shiraishi16, campisi16,polettini17,pietzonka18,luo18,holubec18,benenti20,cangemi20,cangemi21} and how
these are affected by coherence, entanglement, and quantum fluctuations~\cite{scully11,uzdin15, Coherence2, Petruccione, watanabe17,brandner17,carrega19, brandner20,francica20},
and what is the minimum temperature achievable
in a given time in small quantum refrigerators~\cite{levy12,benenti15,paz1,paz2,clivaz19}, are vital for the
construction of quantum machines.
For instance, manipulating heat flows via devices like thermal switches,
diodes,
and transistors is essential to evacuate heat in quantum processors\cite{Pekola21,Giazotto2006,p14}, 
while efficient cooling is strictly related to the preparation of a target
state, say for a qubit, with the desired fidelity and in the shortest
possible time\cite{paz1, miller19, abiuso20, pancotti20}.

Master equations are an invaluable tool to investigate the dynamics of
open quantum system~\cite{breuerbook, weiss}. On the other hand,
such equations typically rely on
the assumptions that the system-baths couplings are weak and the baths 
large enough to neglect non Markovian effects in the
system's dynamics. These approximations, which are quite natural for
macroscopic systems, easily break down
when considering small quantum systems. Such considerations motivate the
huge effort under way in the development of
methods and tools to address regimes of strong coupling and where memory
effects and system-baths quantum correlations
are important~\cite{ciccarello21, nazir14,breuer16,devega17,carrega15, hur14, aurell18, flindt21_2, segal21, campeny21, jurgen, simone}.

Here, we develop a general framework to address the above questions in
periodically driven quantum systems.
In particular, we focus on the work contributions related to the
system-bath couplings.
These terms, overlooked in standard master equation approaches, may play
a relevant role, especially when dealing with cyclic processes, like
in heat engines or refrigerators. Indeed, in one cycle the system is
connected/disconnected to/from one or more baths several times, in a way
dependent on the specific protocol. The work contribution
related to the bath couplings was found to be detrimental to the
efficiency of a four-stroke Otto engine~\cite{jurgen}. This raises the
question, whether this coupling work is always deleterious,
as one might intuitively argue by associating some dissipation to the
non adiabatic switching on/off of the couplings.

In contrast with such naive argument, here we show that, by
periodically modulating only the system-bath couplings in a suitably engineered way,
 it is possible to perform fundamental quantum thermodynamic tasks.
Specifically, we show that it is possible to cool a reservoir at low temperature decreasing the coupling strength, so that the zero
temperature limit can be achieved only in infinite time, in
 agreement with Nernst's unattainability principle~\cite{levy12,benenti15,paz1,paz2}. This surprising
effect is possible due to the asymmetric time-dependent
couplings of the system to two heat baths, and the mixing of different harmonics. We point out that, differently from the ratchets effects discussed in
the literature due to modulated baths' temperatures \cite{hanggi1, hanggi2},
 where heat is \emph{evacuated} asymmetrically in the two reservoirs,
in our case we have a refrigerator and
 therefore heat is \emph{extracted} from the cold reservoir.
 Quite interestingly, this effect takes place only in the low
temperature, non Markovian quantum regime,
 and it vanishes in the high temperature, Markovian classical regime.

 We also show that the same setup, with asymmetrically modulated
couplings {to two baths at temperatures $T_1$ and $T_2$},
can be used to build a dynamically-induced heat rectifier. 
Here large rectification can be obtained also in the
classical regime, {with the possibility to achieve ideal rectification,
blocking of the heat flow in the forward configuration (say, with $T_1>T_2$). 
Conversely, and remarkably, the low-temperature
quantum regime turns out to be more versatile. 
Indeed, by simply tuning the frequency of the couplings modulation,
dynamical heat rectification can be obtained in both directions, eventually blocking the
heat flow in either the forward ($T_1>T_2$) or the backward ($T_1<T_2$) 
configuration}. Interestingly,
this peculiar operating regime corresponds to the one where
refrigeration properties can be obtained.

 The paper is organized as follows. In Sec.~\ref{model_gen} we present the
general setting of periodically driven system-baths couplings and the definitions/expressions of the key thermodynamic quantities. The general approach, based on the out of
equilibrium Green function,  is presented in Sec.~\ref{green}. Here, we solve the dynamics at long times 
and we evaluate the time average (over one cycle) of the heat currents. 
Sec.~\ref{results} is devoted to the discussion of our main results, namely cooling, refrigeration and rectification properties induced by the the time-dependent system-baths coupling. Sec.~\ref{conclusions} contains the summary and conclusions.
Technical details can be found in several Appendices.

\section{General setting}
\label{model_gen}

\subsection{Model}
\label{model}
\begin{figure}
\includegraphics[width=0.9\linewidth]{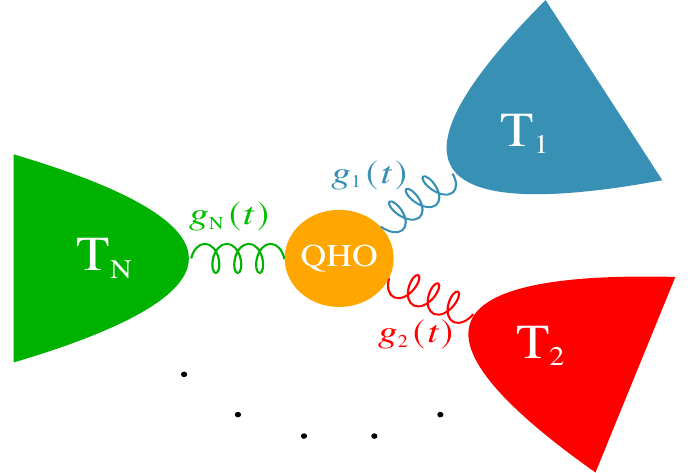}
\caption{\label{fig:0} Sketch of the quantum system coupled to different baths with driven system-baths couplings.
}
\end{figure}

We consider a quantum system linearly coupled to a set  of $N$ reservoirsas sketched in Fig.\ref{fig:0}, described by the total Hamiltonian (hereafter we set $\hbar= k_{{\rm B}}=1$)
\be
\label{eq:hamtot}
H^{(t)}  = H_{{\rm S}} + \sum_{\nu=1}^N \Big[H_\nu + H^{(t)}_{{\rm int},\nu}\Big].
\ee
Each bath is modelled as a collection of harmonic oscillators in the usual Caldeira-Leggett~\cite{weiss, carrega15, aurell18, CL83, zherbe95} framework:
\begin{equation}\label{eq:bathsupp}
H_{{\nu}}=  \sum_{k=1}^{\infty} \qty[\frac{P^2_{k,\nu}}{2 m_{k,\nu}} + \frac{m_{k,\nu} \omega^2_{k,\nu }X^2_{k,\nu}}{2}].
\end{equation}
The interaction parts are  bilinear in the position operators of system and reservoirs, and are assumed time dependent in order to modulate in time their couplings as pictorially shown in Fig.\ref{fig:0}. They read 
\begin{equation}\label{eq:intsupp}
H^{(t)}_{{\rm int,\nu}}= \sum_{k=1}^{\infty}\left\{-x g_\nu(t)c_{k,\nu}X_{k,\nu} +x^2g_{\nu}^2(t)\frac{c^2_{k,\nu}}{2m_{k,\nu} \omega^2_{k,\nu}}\right\},
\end{equation}
where we have introduced, here and in Eq.~\eqref{eq:hamtot}, the superscript index $(t)$ to indicate the parametric time-dependence {for observables}, related to the driven couplings~\cite{footnote:Heisenberg}. In the following, we will focus on cyclic processes, that are governed by  dimensionless and periodic time dependent functions  $g_\nu(t)=g_\nu(t+ {\cal T})$, whose  Fourier expansions read
\be
\label{serieseg}
g_{\nu}(t)=\sum_{n=-\infty}^{+\infty}g_{n,\nu} e^{-in\Omega t},\quad \Omega=\frac{2\pi}{{\cal T}}.
\ee
 The interaction strengths are described by the parameter $c_{k,\nu}$\cite{weiss}, and for this reason 
we consider bounded functions $|g_\nu(t)|\le 1$.

In this work, the system considered is a single quantum harmonic oscillator (QHO), which represents a common building block for several quantum technology platforms\cite{zherbe95, qt1},
\begin{equation}\label{eq:hosc}
H_{{\rm S}}= \frac{p^2}{2m} + \frac{1}{2}m\omega^2_{0} x^2, 
\end{equation}
with $m$ and $\omega_0$ the mass and the characteristic frequency, respectively.

At initial time $t_0$ the baths are assumed in their thermal equilibrium at temperatures $T_\nu$, with the total  density matrix, describing system plus reservoirs,  written in a factorized form as $\rho(t_0) = \rho_{\rm S}(t_0)\otimes \rho_{1}(t_0)\otimes \dots \otimes \rho_{N}(t_0)$, with $\rho_{\rm S}(t_0)$ the initial system density and 
\be
\label{initial}
 \rho_{\nu}(t_0)=\exp(-H_{\nu}/T_\nu)/\Tr{\exp(-H_{\nu}/T_\nu)}
\ee
the thermal density of each bath.

Using the total Hamiltonian~\eqref{eq:hamtot}, we can explicitly write  the equations of motion (EOM)  for the QHO operators $(x(t),p(t))$ and 
for the baths oscillator operators $(X_{k,\nu}(t),P_{k,\nu}(t))$. We have 
\begin{eqnarray}\label{eq:EOM}
\dot{x}(t)&\!\!=\!\!& \frac{p(t)}{m},\nonumber\\
\dot{p}(t)&\!\!=\!\!&-m\omega^2_0 x(t)\nonumber\\
&+&\sum_{\nu=1}^N \sum_{k=1}^{\infty}g_\nu(t) c_{k,\nu} \Big[X_{k,\nu}(t)-\frac{g_\nu(t)c_{k,\nu}}{m_{k,\nu} \omega^2_{k,\nu}} x(t)\Big],\end{eqnarray}
and
\begin{eqnarray}\label{eq:EOM1}
\dot{X}_{{k,\nu}}(t)&=& \frac{P_{k,\nu}(t)}{m_{k,\nu}},\nonumber\\
\dot{P}_{{k,\nu}}(t)&=&- m_{k,\nu}\omega^2_{k,\nu} X_{k,\nu}(t) +g_\nu(t) c_{k,\nu} x(t).
\end{eqnarray}

The solution for the baths degrees of freedom can be written as  a function of the initial conditions and of the operator $x(t)$ as\cite{cangemi21, paz1, zherbe95} 
\beq
\label{EOMbath}
&&{X_{k,\nu}}(t)=X_{k,\nu}(t_0)\cos[\omega_{k,\nu}(t-t_0)]\nonumber\\
&& {+\frac{P_{k,\nu}(t_0)}{m_{k,\nu}\omega_{k,\nu}}\sin[\omega_{k,\nu}(t-t_0)]}\nonumber\\
&& +\frac{c_{k,\nu}}{m_{k,\nu}\omega_{k,\nu}}\int_{t_0}^{t}\!\mathrm{d}s g_\nu(s)x(s) \sin[\omega_{k,\nu}(t-s)].
\eeq
Substituting these expressions into Eq.~\eqref{eq:EOM} one obtains the   generalized quantum Langevin equation for the oscillator
\begin{eqnarray}\label{eq:diffqoper}
&&\ddot{x}(t) + \omega^2_0 x(t)+\int_{t_0}^{+\infty}\!\!\mathrm{d}s\sum_{\nu=1}^N g_\nu(t)\gamma_\nu(t-s)
\nonumber\\
&&
{\times} \Big[\dot{g}_\nu(s)x(s)+
\dot{x}(s)g_\nu(s)\Big] =\frac{1}{m}\sum_{\nu=1}^N g_\nu(t)\xi_\nu(t)~.
\end{eqnarray}
Here,
\begin{equation}\label{eq:gamma}
\gamma_\nu(t)=\frac{\theta(t)}{m}\sum_{k=1}^{\infty}\frac{c_{k,\nu}^2}{m_{k,\nu}\omega_{k,\nu}^2}\cos(\omega_{k,\nu} t)
\end{equation}
represents the memory damping kernel, with $\theta(t)$ the Heaviside step function. Notice that in the rigth hand side of Eq.~(\ref{eq:diffqoper}) we have dropped the inhomogeneous term  $-x(t_0)\sum_{\nu=1}^Ng_\nu(t)g_\nu(t_0)\gamma(t-t_0)$ since it is a typical transient contribution which decays to zero at $t>0$ times {(we assume the initial condition at time $t_0\to-\infty$)}. The operator 
\beq
\label{randomF}
\!\!\xi_\nu(t)&=&\!\!\sum_{k=1}^{\infty}c_{k,\nu}[X_{k,\nu}(t_0)\cos\omega_{k,\nu} (t-t_0)\nonumber\\
&&+\frac{P_{k,\nu}(t_0)}{m_{k,\nu}\omega_{k,\nu}}\sin\omega_{k,\nu} (t-t_0)]
\eeq
is the fluctuating force of the bath $\nu$, and it depends explicitly on the initial conditions of the bath position/momentum operators $X_{k,\nu}(t_0)$ and $P_{k,\nu}(t_0)$.  It has zero average $\langle\xi_\nu(t)\rangle=0$, as one  can see using the initial thermal conditions (\ref{initial}). Notice that here and below we denote the quantum average of any operator $O$ as $\langle O\rangle = {\rm Tr}[O\rho(t_0)]$.

The corresponding correlation functions $\langle\xi_\nu(t)\xi_{\nu'}(t')\rangle$ are evaluated by expressing them in terms of the bath spectral density 
defined as\cite{weiss} \begin{equation}
\label{eq:spectral}
{\cal J}_\nu(\omega)=\frac{\pi}{2}\sum_{k=1}^\infty \frac{c_{k,\nu}^2}{m_{k,\nu}\omega_{k,\nu}}\delta(\omega-\omega_{k,\nu}).
\end{equation}
We have
\be
\label{xicorr}
\langle\xi_\nu(t)\xi_{\nu'}(t')\rangle=\delta_{\nu,\nu'}\left[{\cal L}^{(+)}_\nu(t-t')-i{\cal L}^{(-)}_\nu(t-t')\right],
\ee
where
\beq
\label{correlator}
{\cal L}^{(+)}_\nu(t)&=&\int_0^{\infty}\!\!\frac{\mathrm{d}\omega}{\pi} 
{\cal J}_\nu(\omega)\coth(\frac{\omega}{2T_\nu})\cos(\omega t),\nonumber\\{\cal L}^{(-)}_\nu(t)&=&\int_0^{\infty}\!\!\frac{\mathrm{d}\omega}{\pi} 
{\cal J}_\nu(\omega)\sin(\omega t)
\eeq
describe the symmetric and antisymmetric contributions, respectively.

\noindent Similarly, the damping kernel $\gamma_\nu(t)$ in Eq.~(\ref{eq:gamma}) can be written as 
\begin{equation}\label{eq:gammt}
\gamma_\nu(t)=\frac{2}{\pi m}\theta(t)\int_0^{\infty}\mathrm{d}\omega \frac{{\cal J}_\nu(\omega)}{\omega}\cos(\omega t).
\end{equation}

\subsection{Average thermodynamic quantities}
We now focus on the long time behaviour of the full system (QHO plus baths). In this limit, due to the presence of dissipation, the transient dynamics is  washed out and  the total system reaches a periodic state  sustained by the coupling drives. The key thermodynamic quantities to be inspected are then the driving powers and  the  heat currents averaged over the period ${\cal T}$ of the cycle.
To obtain these quantities we start by defining their  time evolutions $P_\nu(t)$ and $J_\nu(t)$.
The {injected} power $P_\nu(t)$, associated to the temporal variation of the system-bath couplings $g_\nu(t)$, is defined as the following quantum average\cite{cangemi21, carrega19}
\be
\label{powerM}
P_\nu(t)={\rm Tr}\Big[\frac{\partial H_{{\rm int},\nu}^{(t)}}{\partial t}\rho(t)\Big]
\ee
{with $\rho(t)$ the total density matrix at time $t$.}
The corresponding mean power $P_\nu$, averaged over the cycle,  is then given by
\be
P_\nu=\frac{1}{\cal T}\int_0^{{\cal T}} dt\,P_\nu(t),
\ee
with  the total average power $P=\sum_{\nu=1}^N P_\nu$.

The time-dependent heat current, associated to the bath $\nu$,  is given by the change in time of  the reservoir energy. It reads
\be
\label{heatM}
J_\nu(t)=-{\rm Tr}\Big[ H_\nu\dot{\rho}(t)\Big],
\ee
where the minus sign implies a positive $J_{\nu}(t)$ when the energy flows from the reservoir into the quantum system.
The  mean heat current, averaged over the cycle, is then 
\be
J_\nu=\frac{1}{\cal T}\int_0^{{\cal T}} dt J_\nu(t).
\ee

Notice that, in the considered periodic regime, the total  power injected from the coupling drives is balanced by the reservoir heat currents 
and it fulfills  the relation 
\be
 \label{firstlaw1}
P +\sum_{\nu=1}^NJ_\nu=0.
\ee
This equality is derived using the fact that both ${\rm Tr}[H_S{\rho}(t)]$ and ${\rm Tr}[  H_{{\rm int},\nu}^{(t)}{\rho}(t)]$ show a periodic behaviour 
at long times~\cite{jurgen}. 
Eq.~(\ref{firstlaw1}) can be interpreted as the analogue of the first law of thermodynamics. To better see this point we can rewrite it in a more standard form by  introducing the total mean work per cycle $W$ and the mean heat  $Q_\nu $  of the $\nu$ bath\cite{cangemi21, carrega19, jurgen}. These two quantities are directly given by 
$W =P {\cal T}$ and $Q_\nu =J_\nu {\cal T}$ and fulfill   the relation $W+\sum_{\nu=1}^N Q_\nu =0$.

Another key quantity to consider is 
the  time average entropy production rate which is linked to the currents through the relation (see, {e.g.},~\cite{benenti17})
\be
 \label{entropy}
\dot S=-\sum_{\nu=1}^N\frac{J_\nu}{T_\nu}.
\ee
Notice that from the second law of thermodynamics it is  always $\dot S\ge 0$, which in particular implies, for  isothermal baths 
($T_\nu=T$) and using (\ref{firstlaw1}), a positive total power $P=T\dot S\ge 0$. {This relation is consistent with 
Kelvin-Planck  statement of the second law: A cyclic transformation whose sole effect is to convert heat,
extracted from a source at constant temperature, into work, is impossible \cite{cal, kon}.}

We conclude this general part by writing the explicit expressions of the time dependent powers (\ref{powerM}) and heat currents (\ref{heatM}) in terms of the quantum average over the bath and system variables. 
In particular, for the time-dependent heat currents,
inserting the explicit expressions of the Hamiltonian contributions in Eqs.~(\ref{eq:bathsupp}-\ref{eq:intsupp}) one arrives at 
\begin{equation}\label{eq:jsupp}
\!\!\!\!J_\nu(t)=-g_\nu(t)\sum_{k=1}^{\infty}\frac{c_{k,\nu}}{2m_{k,\nu}}\left\langle [x(t) P_{k,\nu}(t) + P_{k,\nu}(t)x(t)]\right\rangle.
\ee
\noindent Now, using the solution of the EOM (\ref{eq:EOM1}) and (\ref{EOMbath}), we have 
\beq
\label{heatexpression}
&&J_{\nu}(t)=-\frac{g_\nu(t)}{2}\left\langle x(t)\dot\xi_\nu(t)+\dot\xi_\nu(t)x(t)\right\rangle \\
&&-g_\nu(t)\int_{-\infty}^{t}\!\!\!\!\!\!\!\mathrm{d}s 
\left\langle x(t)x(s)+x(s)x(t)\right\rangle
g_\nu(s) \frac{{\mathrm d}}{{\mathrm d}t}
{\cal L}^{(-)}_\nu(t-s).\nonumber
\eeq
Following the same steps, the time-dependent power contributions read
\beq
\label{Pexpression}
&&P_\nu(t)=-\dot g_\nu(t)\langle x(t)\xi_\nu(t)\rangle\nonumber\\
&&+m\dot g_\nu(t)\int_{-\infty}^{+\infty}\!\!\!\!\!\!\!\mathrm{d}s\gamma_\nu(t-s)\frac{{\mathrm d}}{{\mathrm d}s}
\left[g_\nu(s)\langle x(t)x(s)\rangle\right].
\eeq
These expressions represent   the key starting point in order to evaluate all thermodynamic quantities after averaging over the period of the cycle.

\section{General approach: out of equilibrium Green function}
\label{green}
To evaluate the above  thermodynamic quantities it is first necessary to solve the EOM of the oscillator position operator $x(t)$ in Eq.~(\ref{eq:diffqoper})\cite{paz1, paz2, zherbe95, arrachea12a}.
To this end, we introduce the  associated retarded Green function $G(t,t')$ which fulfills the following equation:
\begin{eqnarray}
\label{eomgreen}
&&\ddot{G}(t,t') + \omega_0^2 G(t,t')+\int_{-\infty}^{+\infty}\!\!\mathrm{d}s\sum_{\nu=1}^{N}g_\nu(t)\gamma_\nu(t-s)\nonumber\\
&&{\times} [\dot{g}_\nu(s)G(s,t')+
g_\nu(s)\dot{G}(s,t')] =\delta(t-t'),
\end{eqnarray}
with  $G(t,t')=0$ for $t\le t'$. 
Here, the dot denotes the derivative with respect to the first argument. Notice that due to the breaking of temporal translation, caused by the time-dependent couplings, the Green function depends separately on $t,t'$ and not on their difference $t-t'$ only. 
As already mentioned, we are interested in finding solutions  in the long time limit, where the memory of the initial state is lost and 
the system reaches a periodic state substained by the drives. In this regime, the time evolution of the position operator $x(t)$ can be expressed directly as a time integral of the retarded Green function with  the inhomogeneous term:
\be
\label{solution}
x(t)=\sum_{\nu=1}^{N}\int_{-\infty}^{+\infty}\mathrm{d}t' G(t,t')\frac{1}{m}g_\nu(t')\xi_\nu(t').
\ee
This important relation allows us to express all quantum correlation averages, which define the driving powers and the heat currents, in terms of the resolvent Green function $G(t,t')$. 

In addition,  in the long time limit,  the Green function acquires a peculiar time property~\cite{paz1,paz2,arrachea12a, grifoni95, grifoni96, campeny19}. Indeed, even if $G(t,t')$ is not periodic,  the Fourier transform 
\be
\tilde{ G}(t,\omega)=\int_{-\infty}^{+\infty}\!\!\!\!\!\!\!\mathrm{d}t' e^{i\omega(t-t')}G(t,t')
\ee
obeys $\tilde{ G}(t+{\cal T},\omega)={\tilde{ G}}(t,\omega)$ and it can be written in terms of the Fourier series:
\be
\tilde{ G}(t,\omega)=\sum_{m=-\infty}^{+\infty}\tilde{ G}_m(\omega) e^{-im\Omega t},
\ee
with  $\tilde{ G}_m(\omega)$  the so-called  Floquet coefficients.
Notice that for static drive, only the $m=0$ component  would be present in the series expansion. Using the above relation we can write a rather compact expression for $G(t,t')$:
\be
\label{transform}
G(t,t')=\sum_{m=-\infty}^{+\infty}\int_{-\infty}^{+\infty}\frac{\mathrm{d}\omega}{2\pi}e^{-i\omega(t-t')}\tilde{ G}_m(\omega) e^{-im\Omega t}.\ee
Inserting (\ref{transform}) into 
Eq.~(\ref{eomgreen}) one obtains the following algebraic set of equations for the Floquet coefficients (see Appendix \ref{Floquet} for more details):
\beq
\label{floquet1}
\tilde{ G}_m(\omega)&=&\chi_0(\omega)\delta_{m,0}-\chi_0(\omega+m\Omega)\nonumber\\
&&{\times}\sum_{n\neq 0}\tilde{k}_n(\omega+(m-n)\Omega) \tilde{ G}_{m-n}(\omega), 
\eeq
where
\be
\label{kappan}
\tilde{k}_n(\omega)=-i\sum_{\nu=1}^N \sum_{m=-\infty}^{+\infty}g_{m,\nu}g_{n-m,\nu}(\omega+m\Omega)\widetilde\gamma_\nu(\omega+m\Omega)
\ee
represents the influence kernel of the driving due to the baths, with $\widetilde\gamma_\nu(\omega)$ the Fourier transform of $\gamma_\nu(t)$ in Eq.~(\ref{eq:gammt}).
In Eq.~(\ref{floquet1}), we have also introduced the ``static'' retarded Green function in spectral domain
\be
\label{chi}
{\chi}_0(\omega)=-\frac{1}{\omega^2-\omega_0^2-\tilde{k}_0(\omega)},
\ee
which contains the $n=0$ component, $\tilde{k}_0(\omega)$, of the bath kernel $\tilde{k}_n(\omega)$. 

The important point of the present approach is that the knowledge of the Floquet coefficients $\tilde{ G}_m(\omega)$ allows to solve the full dynamics of the system at long times. 
We underline that  the   solution of the coupled equations (\ref{floquet1})  in general should be tackled by means of numerical techniques, i.e. by exploiting exact diagonalization and inversion of large matrices or using an iterative procedure. 

We conclude this general part by deriving the expressions for the average heat currents and powers  written in terms of the Floquet coefficients.
Here, we quote the main steps, presenting details in Appendix \ref{heatpowerapp}. First of all, we consider the expression for the position operator in Eq.~(\ref{solution}) and we insert it  into the  heat current $J_{\nu}(t)$ in Eq.~(\ref{heatexpression}). The result is 
\beq
\label{appja}
&&J_{\nu}(t)= -\frac{g_\nu(t)}{m}\int_{-\infty}^{+\infty}\!\!\!\!\!\!\!\mathrm{d}t' G(t,t')g_{\nu}(t')\dot{\cal L}_\nu^{(+)}(t-t')\nonumber\\&& -\frac{2g_\nu(t)}{m^2}{\int_{-\infty}^{t}}\!\!\!\!\mathrm{d}s\int_{-\infty}^{+\infty}\!\!\!\!\!\!\mathrm{d}t_1\int_{-\infty}^{+\infty}\!\!\!\!\!\!\!\mathrm{d}t_2\, g_\nu(s)\dot{\cal L}_\nu^{(-)}(t-s)\nonumber\\
&&{\times} G(t,t_1)G(s,t_2)\sum_{\nu_1=1}^{N}g_{\nu_1}(t_1)g_{\nu_1}(t_2){\cal L}_{\nu_1}^{(+)}(t_1-t_2),
\eeq
where, as before, the  dot denotes the derivative with respect to the first argument. Using now the Fourier integrals and series of  Eqs.~(\ref{serieseg}),(\ref{transform}) we obtain  the heat current after the time average over the cycle: 
\beq
\label{jaaverage}
&&J_\nu=\sum_{n_1,n_2=-\infty}^{+\infty}\!\!\!\!\!\!
g_{n_1,\nu}g_{n_2,\nu}\int_{-\infty}^{+\infty}\!\!\frac{\mathrm{d}\omega}{2\pi m}\Big\{-i {\cal J}_\nu(\omega)\omega\nonumber\\&&
{\times} \coth(\frac{\omega}{2T_\nu}) \tilde{ G}_{-(n_1+n_2)}(\omega+n_2\Omega)
\nonumber\\
&&
-\sum_{\nu_1=1}^{N} \sum_{m_1=-\infty}^{+\infty}
\sum_{n_3, n_4=-\infty}^{+\infty}\!g_{n_3,\nu_1}g_{n_4,\nu_1}\!\frac{{\cal J}_{\nu_1}(\omega)}{m}\coth(\frac{\omega}{2T_{\nu_1}})\nonumber
\\&&\nonumber\\
&&{\times} [\omega-\Omega(n_2+n_4+m_1)]{\cal J}_{\nu}(\omega-\Omega(n_2+n_4+m_1))\nonumber\\&&\nonumber\\
&&{\times} \tilde{ G}_{m_1}(-\omega+n_4\Omega)\tilde{ G}_{-(n_{\rm tot}+m_1)}(\omega+n_3\Omega)\Big\},
\eeq
with $n_{\rm tot}=n_1+n_2+n_3+n_4$. In the above expression the spectral density ${\cal J}_\nu(\omega)$ is extended at negative frequencies by requiring the odd property 
${\cal J}_\nu(\omega)=-{\cal J}_\nu(-\omega)$.

Starting from Eq.~(\ref{Pexpression}) and following similar steps we obtain also the average power associated to the $\nu$-th bath (see Eq.(\ref{Paverage}) in Appendix \ref{heatpowerapp}). 

\section{Results and discussion}
\label{results}
The formalism developed so far is general and allows to exactly evaluate all stationary heat currents and power contributions in a multiterminal configuration with $\nu$ reservoirs in presence of time-dependent drives modulating the various system-bath couplings. 

Hereafter,  we will consider Ohmic baths, that describe a wide class of dissipative environments, with spectral densities 
\be
\label{j_ohm}
{\cal J}_\nu(\omega)=m\gamma_\nu
\omega e^{-|\omega|/\omega_c},
\ee
where $\gamma_\nu$ quantifies the interaction strength between the $\nu$-th bath and the system, and $\omega_c$ is the bath  cut-off frequency kept as the largest energy scale. This important class of dissipation has damping kernels $\gamma_\nu(t)$ local in time, $\gamma_\nu(t)=2\gamma_\nu\theta(t) \delta(t)$,
with Fourier transform
\be
\widetilde\gamma_\nu(\omega)=\gamma_\nu.
\ee
 In this case, the dissipative kernel $\tilde{k}_n(\omega)$  in Eq.~(\ref{kappan}) can be rewritten as the following  time average:  
\be
\label{kappan1}
\tilde{k}_n(\omega)=\frac{1}{{\cal T}}\int_0^{{\cal T}}{\rm  d}t e^{in\Omega t}\sum^N_{\nu=1}\gamma_\nu[g_\nu(t)\dot g_\nu(t)-i\omega g_\nu^2(t)].
\ee
It is worth stressing that, although the damping kernels $\gamma_\nu(t)$ are local in time with Ohmic spectral functions, this does not imply a Markovian dynamics. Indeed, the noise terms in Eq.~\eqref{randomF}, retain memory and thus non Markovian signatures. It is only in the classic regime, {$T\gg \omega_0$}, that  noise terms loose memory, as one can see looking at the correlators~\eqref{xicorr} which become
\be
\label{lim_classic_corr}
 \langle \xi_\nu(t)\xi_{\nu'}(t')\rangle \approx 2m\gamma_\nu T_\nu \delta_{\nu,\nu'}\delta(t-t').
\ee

\subsection{Ratchet induced cooling and refrigeration}
We now discuss the average heat flows induced by the temporal modulation of the system-bath couplings. The aim is to find particular regimes where time-dependent coupling act in a thermodynamical efficient way, opposite to the usually expected {purely} dissipative regime. 

To this end, we will consider peculiar shapes of the drives $g_\nu(t)$, that break time-translational invariance and can allow for the so-called heat ratchet effect. It has been shown that asymmetric drives, by relying on dynamical symmetry breaking, are indeed able to produce e.g. directed heat flow and heat rectification~\cite{hanggi1,hanggi2, hanggi3}. Dynamical breaking of temporal reflection symmetry, and thus ratchet effect, can be induced by {nonlinear} harmonic mixing of different frequencies {of the drives. Note that such mixing is possible, in spite of 
the fact we are considering a harmonic oscillator system, due to the periodic drive.}
Importantly, and differently from previous studies, {in our case the ratchet effect is achieved
by suitably engineering the modulation of the system-bath
couplings only.}
To proceed further, we notice that the expression in Eq.~\eqref{kappan1} is quite illuminating, since it allows to find a particular class of time dependent couplings $g_\nu(t)$, which verify  $\sum_{\nu=1}^{N}\gamma_\nu{\dot g}_\nu(t) g_\nu(t)=0$. This implies the following {constraint}: 
\be
\label{vincolo}
\sum_{\nu=1}^{N}\gamma_\nu g_\nu^2(t)={\gamma},
\ee
with $\gamma$ an effective damping that feels both dissipation amplitudes $\gamma_\nu$ and the associated harmonic components.
Notably, the above {constraint} implies that only the static contribution $\tilde{k}_0(\omega)$ in Eq.~(\ref{kappan1}) is different from zero, with
\be
\label{kappa0}
\tilde{k}_n(\omega)=-i\omega\gamma\,\delta_{n,0}.
\ee
The above relation allows an {\it exact} solution of the set of coupled algebraic equations in Eq.~\eqref{floquet1}. Indeed, one finds
\be
\label{GG0}
\tilde{ G}_m(\omega)=\chi_0(\omega)\delta_{m,0},
\ee
with only the static retarded Green function component, even if the couplings $g_\nu(t)$ are still time-dependent. The resulting expressions for the average heat currents are now simplified by putting Eq.~\eqref{GG0} into Eq.~\eqref{jaaverage}. We have 
\beq
\label{jaaveragebis}
&&J_\nu=\gamma_\nu \int_{-\infty}^{+\infty}\!\!\frac{\mathrm{d}\omega}{2\pi}\Big\{\omega^2\coth(\frac{\omega}{2T_\nu}) e^{-|\omega|/\omega_c}
\sum_{n_1=-\infty}^{+\infty}\!\!\!|g_{n_1,\nu}|^2\nonumber\\
&&{\times}\Im\chi_0(\omega-n_1\Omega)-\!\!\!\!
\sum_{\nu_1=1}^{N}\sum_{n_{1,3,4}=-\infty}^{+\infty}\!\!\!\!\!g_{n_1,\nu}g_{-(n_1+n_3+n_4),\nu}\nonumber\\&&
{\times}g_{n_3,\nu_1}g_{n_4,\nu_1}
\!\gamma_{\nu_1}\omega\coth(\frac{\omega}{2T_{\nu_1}})[\omega+\Omega(n_1+n_3)]^2e^{-|\omega|/\omega_c}\nonumber\\&&\nonumber\\
&&{\times}\chi_{0}(-\omega+n_4\Omega)\chi_0(\omega+n_3\Omega)e^{-|\omega+\Omega(n_1+n_3)|/\omega_c}\Big\}.
\eeq

Several choices of $g_\nu(t)$ can fulfill the constraint~\eqref{vincolo}, and we believe that this is a convenient setting to grasp the physics that can be induced by time-dependent system-bath couplings. As an illustrative example we now focus on two reservoirs  $\nu=1,2$ only. Despite its simplicity, it represents {the prototypical model} in order  to investigate the role of driven system-bath coupling on thermodynamic performances.

First of all, we note that monochromatic drives such as $g_1(t)=\sqrt{\gamma/\gamma_1}\cos(\Omega t)$ and $g_2(t)=\sqrt{\gamma/\gamma_2}\sin(\Omega t)$, satisfy the constraint \eqref{vincolo}, but they will not produce any harmonic mixing, thus resulting in a rather trivial dynamics. {For instance}, in such a case, for isothermal baths the average heat currents are always dissipated into the reservoirs, i.e. both $J_1$ and $J_2$ have negative signs, independently of the operating regime (driving frequency or temperature range). 
Therefore, {in order to observe nontrivial effects} we choose
\beq
\label{scelta_vincolo}
g_1(t)&=&\cos(\Omega t),\nonumber\\
g_2(t)&=&\sqrt{\frac{\gamma}{\gamma_2}}\sqrt{1-\kappa\cos^2(\Omega t)}~,
\eeq
where we have defined the effective asymmetry $0< \kappa=\gamma_1/\gamma < 1$. 
This choice allows harmonic mixing between the two drives, as one can see by using Eq.~\eqref{serieseg} and evaluating the associated Fourier coefficients. Indeed, one obtains
\beq
\label{fft_g}
\!\!\!\!&& g_{n,1}=\frac{1}{2}\big[\delta_{n,1} + \delta_{n,-1}\big]\nonumber \\
\!\!\!\!&&g_{n,2}=\sqrt{\frac{\gamma}{\gamma_2}}\Big[ \frac{2}{\pi}{ E}(\kappa) \delta_{n,0} -\frac{\kappa}{8}\, _2F_1(\frac{1}{2},\frac{3}{2},3;\kappa) \delta_{n,\pm2}\nonumber\\
\!\!\!\!&&\!\!-\frac{(n-3)!!}{(n/2)! 2^{3 n/2}}\kappa^{n/2}\!_2F_1\!(\frac{n-1}{2},\frac{n+1}{2},n+1;\kappa) \delta_{n,\pm 2p}\Big]
\eeq
where ${ E}(x)$, $_2 F_1(a,b,c;x)$ are the Ellyptic and Hypergeometric functions, respectively, and $p$ an integer with $p\geq 2$. 
Depending on the value of $\kappa$ an interplay between odd and even harmonics is therefore expected (see Eq.~\eqref{jaaveragebis}).
\begin{figure}
\includegraphics[width=0.95\linewidth]{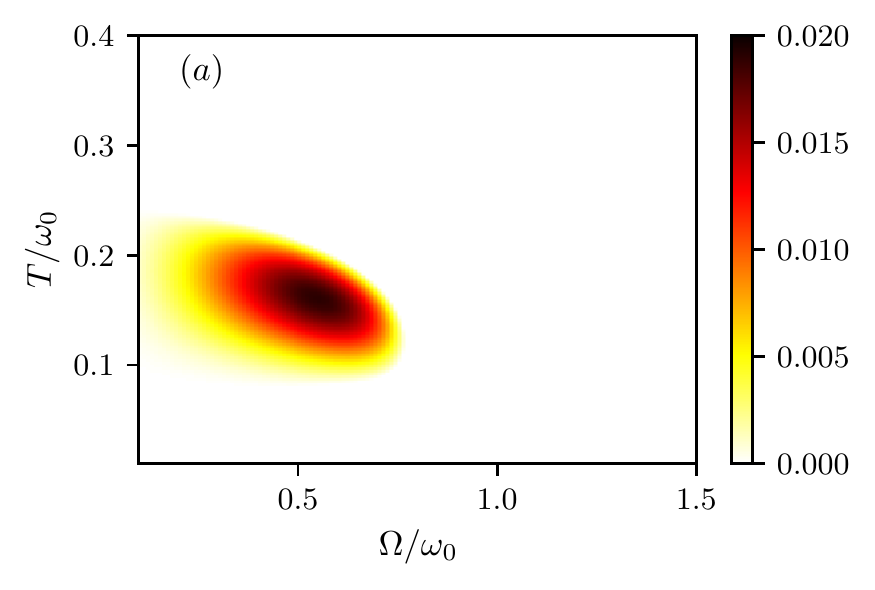}
\includegraphics[width=0.95\linewidth]{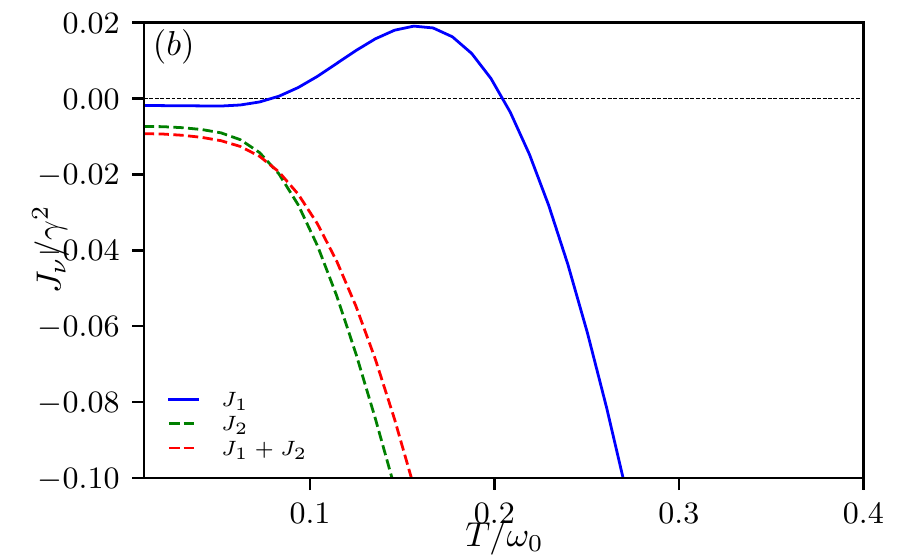}
\caption{\label{fig:1}
Average heat currents with asymmetric coupling $\kappa=\gamma_1/\gamma=0.2$ and isothermal baths $T_1=T_2=T$. Panel (a): density plot of the average heat current ${J}_1/\gamma^2$ as a function of driving frequency $\Omega/\omega_0$ and temperature $T/\omega_0$. Colored regions refer to positive values of $J_1/\gamma^2$ (heat flow from the bath $\nu=1$ towards the system). Panel (b): average heat currents $J_1/\gamma^2$, $J_2/\gamma^2$, and $(J_1+J_2)/\gamma^2$ as a function of temperature $T/\omega_0$ at frequency $\Omega=0.55\omega_0$. Damping strength is $\gamma=0.01\omega_0$ and $\omega_c=500\omega_0$.
}
\end{figure}

We have evaluated the average heat currents $J_\nu$ by means of numerical integration of Eq.~\eqref{jaaveragebis}, using  Eq.~\eqref{fft_g} for different values of $\kappa$. 

An illuminating example is shown  in Fig.~\ref{fig:1} where we consider isothermal baths and a representative value for the effective damping strength and asymmetry. Here, finite heat currents are obtained akin to a {\it dynamical} Peltier contribution. In Fig.~\ref{fig:1}(a) we show a density plot of the average heat current $J_1$ versus frequency $\Omega/\omega_0$ and temperature $T/\omega_0$. Note that for the specific choice of the drives in Eq.~\eqref{scelta_vincolo}, the $\nu=1$ reservoir is the one 
that may exhibit non trivial phenomena. 
This can be seen from the density plot, where positive values $J_1>0$ appear (see the colored area).
This means the presence of an induced {\it ratchet  cooling phenomenon}, with  a heat current that flows from the  $\nu=1$ reservoir towards the system. This behaviour is {counterintuitive, since in the absence of additional external driving forces 
acting on the system, one would naively expect  a dissipative heat current induced by the 
modulation of the system-baths couplings, flowing from the system to the reservoir}. 

It is important to underline that these spots of positive $J_1$ are  present only in the deep quantum regime $T\ll \omega_0$ where non Markovian contributions are present. Indeed, in the classical regime, where the whole system looses memory (Markovian dynamics) 
(see Eq.~\eqref{lim_classic_corr}), both average heat currents, resulting from dynamical Peltier contribution, have always a negative sign (see Appendix \ref{classic} {for a rigorous proof of this result}).

In Fig.~\ref{fig:1}(b),  a cut of the density plot at fixed driving frequency $\Omega$ is shown.  Since the region with positive $J_1$ is always found at frequencies $\Omega \leq \omega_0$ we choose here $\Omega=0.55\omega_0$. In addition to $J_1$,  we plot both the average heat current $J_2$, flowing in the other reservoir $\nu=2$, and the sum of the two heat currents $J_1+J_2$. We underline that the two latter quantities are always negative in the explored parameter regions. This confirms a total power $P=-({J}_1+{J}_2)$, supplied by the external coupling drives,  always positive, in agreement with the relation $P=T\dot S\ge 0$ (see Eq.~(\ref{entropy})).

Importantly, Figure~\ref{fig:1} shows that a non trivial cooling mechanism can emerge, with $J_1>0$, without requiring any external forces directly coupled to the quantum system. To better investigate this phenomenon induced by  time-dependent drives acting on the system-bath couplings, we have studied the behaviour of $J_1$ for different values of the asymmetry parameter $\kappa$. Indeed by varying $\kappa$, one can change the asymmetry between the couplings and, at the same time, increase/decrease the mixing of different harmonics of the time-dependent signals.
As a general result, we observe qualitatively similar behaviours as the ones presented in Fig. \ref{fig:1}. Common ingredients, to obtain the cooling phenomenon, are: 
a ratchet dynamics, a non Markovian behaviour, present only in the deep quantum regime $T\ll \omega_0$, and driving frequencies $\Omega<\omega_0$. Concerning the last inequality, we can say that the stronger is the asymmetry (smaller values of  $\kappa$), the closer is the frequency to resonance ($\Omega\to \omega_0^{-}$) in the region with positive $J_1 >0$.

In Fig.~\ref{fig:2}(a), we plot the maximum positive value of $J_1/\gamma^2$ as a function of the asymmetry parameter $\kappa$. Here, a non monotonic behaviour is visible, 
starting linearly at $\kappa\to 0$ with a maximum around  $\kappa\sim 0.2$. {In Fig.~\ref{fig:2}(b),  the role of $\kappa$ is inspected 
by means of a figure of merit defined as} the ratio between the maximum value  of the average heat current ${\rm Max}[J_1]$ and the corresponding total amount of power $P$ supplied by the drives. In this case, a decreasing behaviour is present and it clearly emerges that $\kappa \ll 1$ is the optimal choice: the normalized cooling effect is higher with stronger asymmetry and it tends to saturate for sufficiently low values of $\kappa$.
Notice that a similar, monotonic and decreasing, behaviour is found also if one considers the maximum of the ratio ${\rm Max}[J_1/P]$ as a function of the $\kappa$, as depicted in the inset to Fig.~\ref{fig:2}(b).
  \begin{figure}
\includegraphics[width=0.95\linewidth]{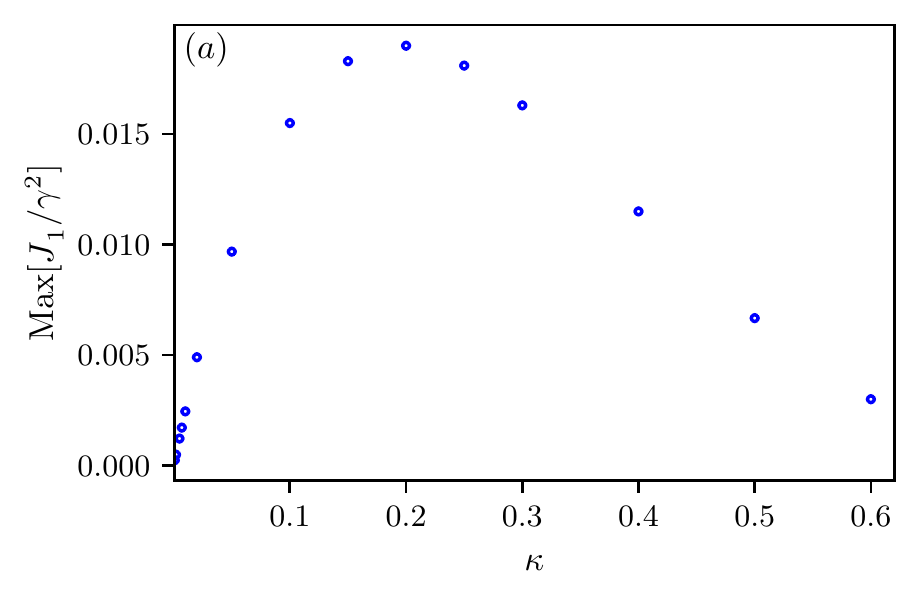}
\includegraphics[width=0.95\linewidth]{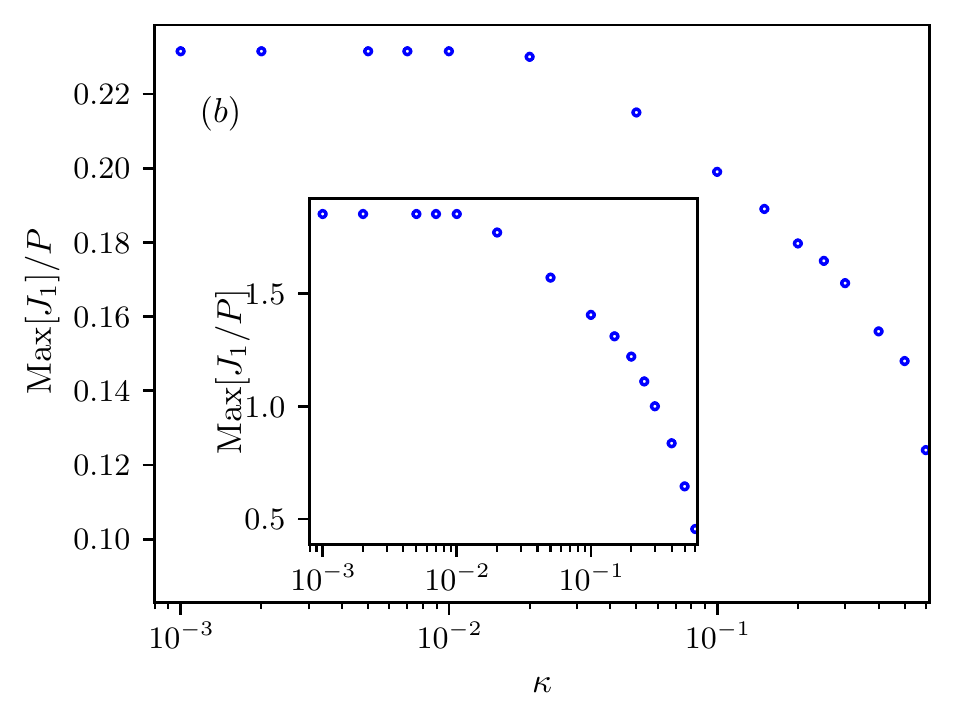}
\caption{\label{fig:2}
Asymmetry dependence of cooling. Panel (a): maximum positive value of ${J}_1/\gamma^2$,  achieved in the $(\Omega, T)$ plane, as a function of the asymmetry parameter $\kappa$. Panel (b): ratio between the maximum positive value of ${J}_1$ and the corresponding average total power $P$ supplied by the drives as a function of $\kappa$. The inset shows the monotonic behaviour also of the maximum of the ratio ${\rm Max}[J_1/P]$ as a function of $\kappa$. Other parameters as in Fig.~\ref{fig:1}(a).
}
 \end{figure}
 
Below, we therefore focus on the strongly asymmetric case $\kappa\ll 1$. Here, a perturbative expansion can be carried out by approximating  $g_\nu(t)$ in Eq.~(\ref{scelta_vincolo}) as
\beq
\label{gpert}
g_1(t)&=&\cos(\Omega t),\nonumber\\
g_2(t)&\approx&\sqrt{\frac{\gamma}{\gamma_2}}[1-\frac{\kappa}{2}\cos^2(\Omega t)].
\eeq
It is worth noting that within this perturbative expansion, up to linear order in $\kappa$, the constraint in Eq.~\eqref{vincolo} is no more guaranteed and one should carefully check that Eq.~\eqref{GG0} is still satisfied at the appropriate truncation of the expansion. This is indeed the case: we have verified (see  Appendix~\ref{complex}) that all other contributions start at order $O(\kappa^2)$. Consistently with such a perturbative approach, the corresponding average heat currents are evaluated  up to linear order in $\kappa$.

The average heat current, associated to the $\nu=1$ reservoir, is (see Appendix~\ref{complex} for details)
\beq
\label{int1}
&&J_{1}=\kappa\gamma \ \int_{-\infty}^{+\infty}\!\!\frac{\mathrm{d}\omega}{4\pi}\,\Big\{
-(\omega^2+\Omega^2)\Im\chi_{0}(\omega)\coth(\frac{\omega}{2T_{2}})\nonumber\\
&&+\omega^2 \Im\chi_0(\omega+\Omega)\coth(\frac{\omega}{2T_1})\Big\}.
\eeq
Similarly the one  for the reservoir $\nu=2$ reads 
\beq
\label{int2}
&&{{J}}_{2}=\kappa\gamma \int_{-\infty}^{+\infty}\!\!\frac{\mathrm{d}\omega}{4\pi}\,\Big\{
\omega^2\Im\chi_{0}(\omega)\coth(\frac{\omega}{2T_{2}})\nonumber\\
&&-\omega(\omega+\Omega)\Im\chi_0(\omega+\Omega) \coth(\frac{\omega}{2T_1})\Big\}.
\eeq
These expressions are well-behaved and therefore we have safely taken the $\omega_c\to\infty$ limit for the cut-off of the Ohmic spectral functions.
Finally, the average total power is obtained from $P=-(J_1+J_2)$.

The above expressions can be analitically evaluated in closed form by resorting to proper Matsubara resummation and integration in the complex plane. Details and full expressions can be found in Appendix~\ref{complex}.
\begin{figure}
\includegraphics[width=0.95\linewidth]{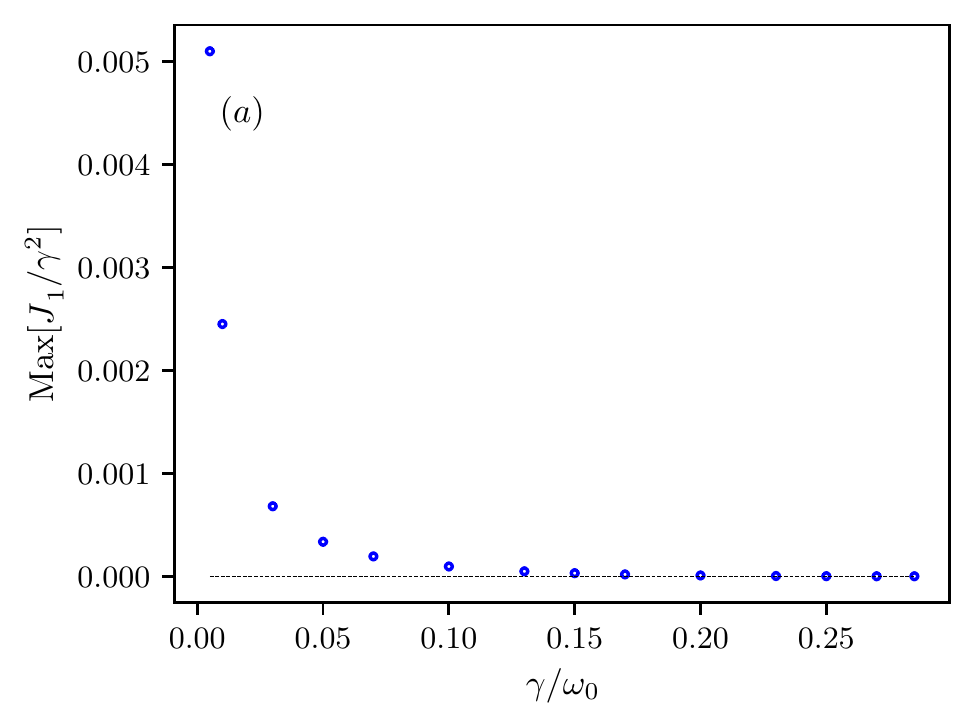}
\includegraphics[width=0.95\linewidth]{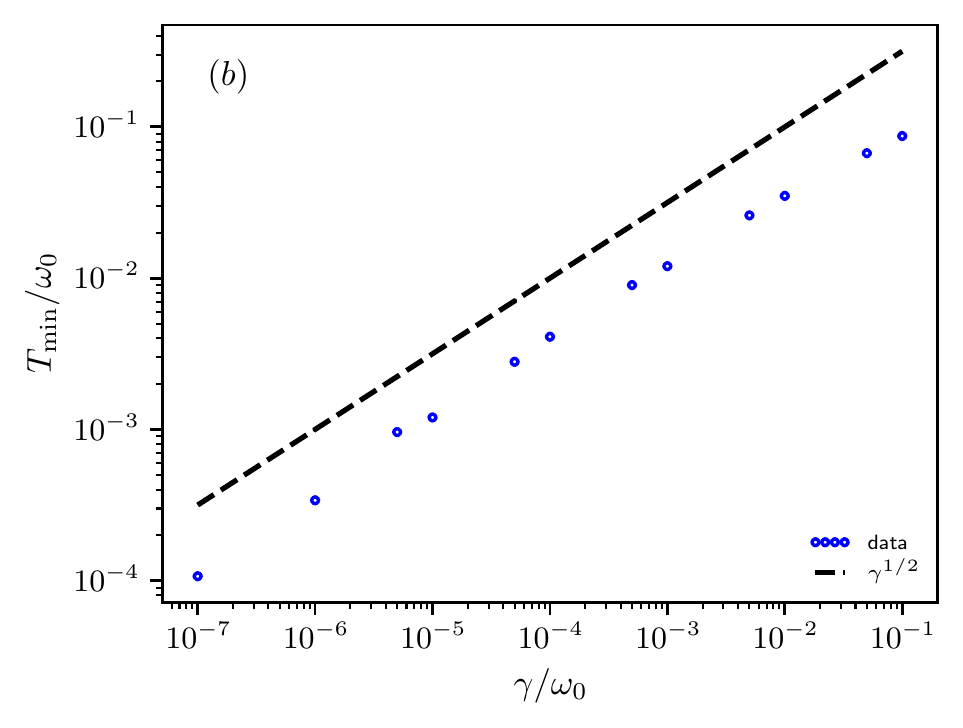}
\caption{\label{fig:3} {Cooling power and minimum achievable temperature versus effective damping.} Panel (a): maximum positive value of $J_1/\gamma^2$ as a function of the effective damping $\gamma$ evaluated in the perturbative regime with strong asymmetry $\kappa=0.01$. Above a critical value $\gamma_c=0.285\omega_0$ no positive values for $J_1$ are found. Panel (b): scaling behaviour of the minimum achievable temperature $T_{{\rm min}}$ with the cooling mechanism as a function of $\gamma$. The data points well agree with the scaling behaviour $\propto \sqrt{\gamma}$. 
}
\end{figure}

We now discuss the cooling properties by inspecting the behaviour of the average heat currents in Eqs.~\eqref{int1}-\eqref{int2}. We start by pointing out that the possibility to achieve regions with positive $J_1$ depends also on the value of the effective damping $\gamma$. Indeed, if $\gamma$ is too strong all heat currents $J_\nu$ dissipate into the reservoirs, with $J_\nu<0$. To elucidate this point, in Fig.~\ref{fig:3}(a) we have depicted the maximum positive value of $J_1$, achieved in the $(\Omega, T)$-plane, as a function of the effective damping $\gamma$ using Eq.~(\ref{int1}) valid for strong asymmetry (in the figure, $\kappa=0.01$). Here, for  $\gamma\to 0$ we have $J_1\propto\gamma$, while increasing $\gamma$ above a critical value $\gamma_c$, $J_1$ becomes negative in the whole parameter range. For the specific value of $\kappa=0.01$, we obtain $\gamma_c=0.285\omega_0$. A qualitatively similar behaviour is also found for other values of $\kappa$, outside the perturbative regime, only with small changes in the precise value of $\gamma_c$.

It is important to stress that, even if there are regions at fixed temperature with $J_1>0$, decreasing the temperature towards $T\to 0$, the average heat current ${J}_1 $ becomes always negative, consistently with the Nernst's unattainability principle \cite{paz1, paz2}. 
To be more quantitative, we look in the above perturbative regime for the zeroes of $J_1 (\Omega, T)$  at a given  effective damping strength $\gamma$. These are points in the $(\Omega, T)$-plane. We define  the minimum achievable cooling temperature $T_{{\rm min}}$ as the one associated to the zero point with the minimum temperature among all the possible zeros of $J_1$. By varying the effective damping $\gamma$ we reproduce the function $T_{{\rm min}}(\gamma)$ shown in Fig.~3(b). As we can see, the  weaker is the damping ({and consequently the cooling power}) the lower is the cooling temperature $T_{{\rm min}}$, 
{so that the $T\to 0$ limit can only be achieved for $\gamma\to 0$ in infinite time, in agreement with 
 Nernst's principle.}
In addition (data not shown here) the corresponding driving frequency is approaching (from below) $\Omega\to \omega_0$. From the plot it is also evident  that the scaling behaviour with $\gamma$ exhibits a $\sqrt{\gamma}$ dependence. 
Notice that a similar scaling was found by Freitas et al.\cite{paz1} in a different context: static system-bath Ohmic coupling and in the presence of an external parametric drive of the oscillator frequency.  We remark that the underlying mechanism discussed here is different, since it relies on temporal variation of the system-bath couplings and ratchet effect without any external field.\\ 
\begin{figure}
\includegraphics[width=0.95\linewidth]{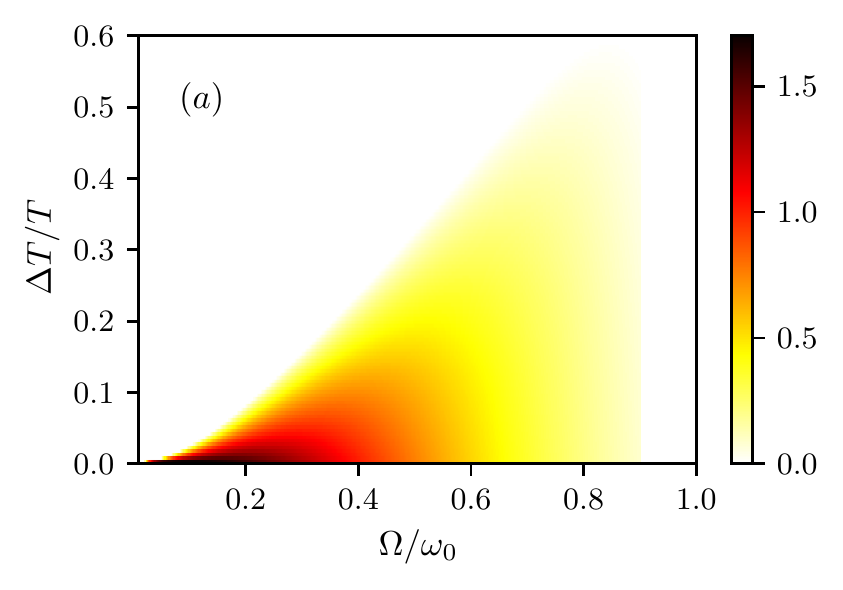}
\includegraphics[width=0.95\linewidth]{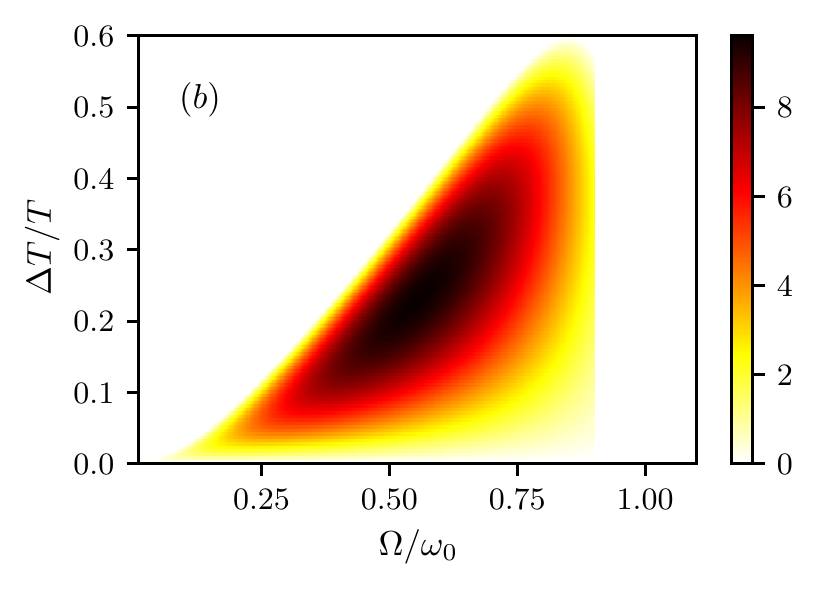}
\caption{\label{fig:4} Ratchet induced refrigeration. Panel (a): density plot of $\eta={J}_1/|{J}_1+{J}_2|$ as a function of $\Omega/\omega_0$ and relative temperature variation $\Delta T/T$ with average temperature $T=0.107\omega_0$. Panel (b): density plot of the ratio $\eta/\eta_C$ (in percentage). Other parameters are $\gamma=0.01\omega_0$ and $\kappa=0.01$.
}
\end{figure}

To further exploit the ratchet induced cooling mechanism discussed until now  for isothermal reservoirs, we consider the two reservoirs at different temperatures. In particular, we choose symmetric deviations from the isothermal situation with $T_{1,2}=T \mp \Delta T/2$, where $T$ represents the average temperature and $\Delta T$ the temperature gradient. Here, in the presence of finite thermal gradient $\Delta T\neq 0$, one can look for refrigeration property\cite{benenti17, r1, r2, vischi19, sing20, bhandari21a}, i.e. heat  extracted from the colder reservoir thanks to the ratchet dynamics induced by the coupling drives.  To quantify this effect we consider the following figure of merit:
\be
\label{eta}
\eta(\Omega,\Delta T/T)= \frac{J_1(\Omega,\Delta T/T)}{|J_1(\Omega,\Delta T/T)+J_2(\Omega,\Delta T/T)|},
\ee
which represents the so-called coefficient of performance (COP) of refrigerators at a fixed average temperature $T$.
This quantity is plotted  in Fig.~\ref{fig:4}(a) as a function of external frequency $\Omega$ and relative temperature gradient $\Delta T/T$. As average temperature we have chosen $T=0.107\omega_0$, that is the one that maximize the $\eta(\Omega,\Delta T=0)$, i.e. the ratio $J_1/P$ in the isothermal case.
 As one can see, the colder $\nu=1$ reservoir, can be cooled ($J_1>0$) in a relatively large parameter region of the  $(\Omega-\Delta T/T)$-plane (see colored area).

In order to quantify the efficiency 
we plotted, in Figure~\ref{fig:4}(b),  the function $\eta$ normalized to the Carnot value for refrigerators\cite{benenti17}:
\be
\label{carnot}
\eta_C={ \frac{T_1}{T_2-T_1}}=-\frac{1}{2} +\frac{T}{\Delta T}~.
\ee
Here, it is possible to achieve value of the COP up to $\sim 10\%$ of $\eta_C$. Remarkably, such values are obtained both in the non adiabatic and non linear regime. In passing, we mention that even higher values of this ratio can be achieved decreasing the effective damping $\gamma$, although the magnitude of the heat currents will be smaller.

\subsection{Beyond dynamical constraints}
\begin{figure}
\includegraphics[width=0.95\linewidth]{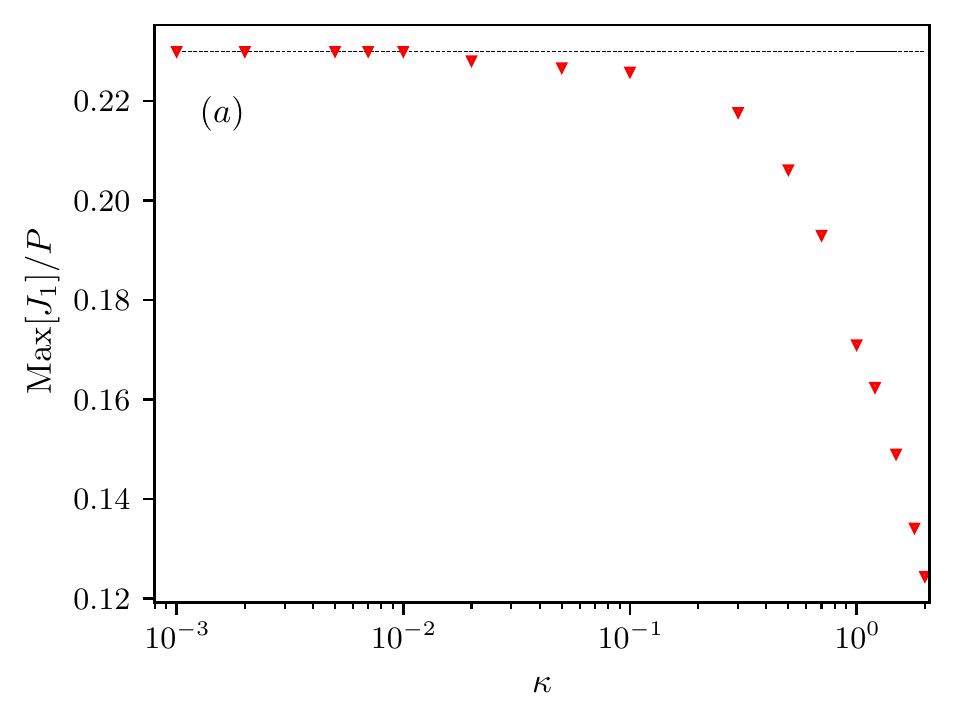}
\includegraphics[width=0.95\linewidth]{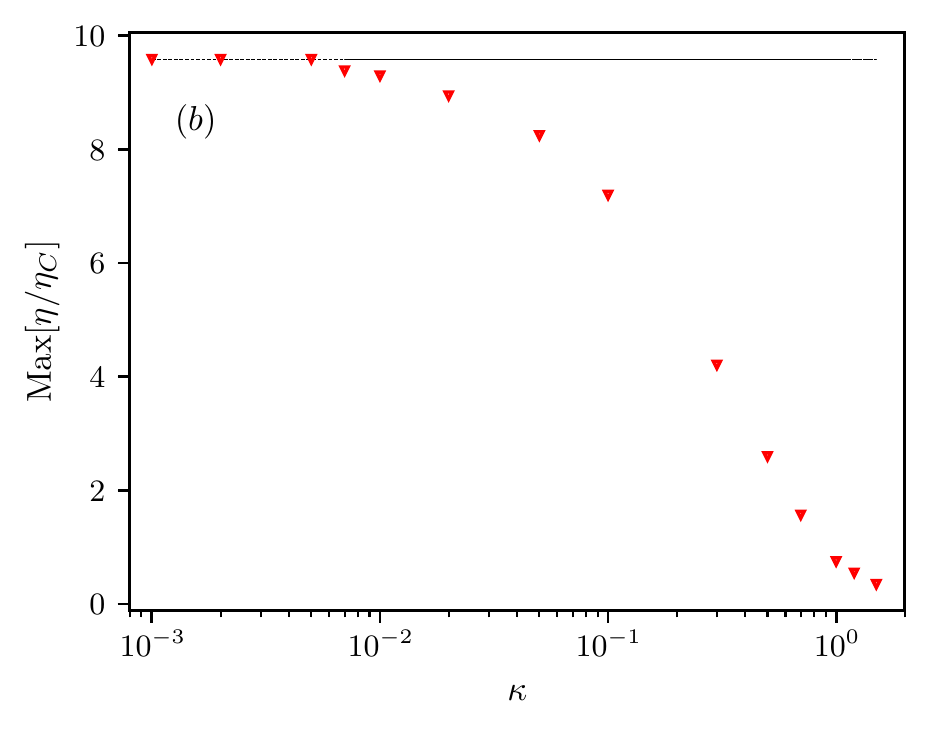}
\caption{\label{fig:5} Ratchet-induced cooling 
beyond the dynamical constraint. Panel (a): maximum positive heat current normalized to the corresponding supplied power ${\rm Max}[J_1]/P$ as a function of $\kappa=\gamma_1/\gamma_2$ in the isothermal case $T_1=T_2$. Panel (b): maximum value of the COP normalized to the Carnot bound ${\rm Max}[\eta/\eta_C]$ in percentage, as a function of $\kappa$ considering finite temperature gradient $\Delta T\neq 0$. The black dashed lines correspond to the asymptotic value obtained within the perturbative expansion. Notice that the data points tend to saturate to the perturbative results for small $\kappa$ values. Other parameters are $\gamma={\rm Max}[\gamma_1,\gamma_2]=0.01\omega_0$ and $\omega_c=500\omega_0$. 
}\end{figure}
The results discussed so far have been obtained within the particular choice~\eqref{vincolo} for the drives, which allows for an exact solution of the problem. At a first sight this could seem a very peculiar case. However, in the following we will demonstrate that the possibility of performing fundamental quantum thermodynamic task is more general and it goes beyond the above assumption. In order to corroborate this point and to verify the stability of the dynamically induced cooling phenomenon we consider different time-dependent drives outside the constraint class. Among all possibilities, as an illustrative example,  we choose
\be
\label{g_num}
g_1(t)=\cos(\Omega t),\quad \quad g_2(t)=1.
\ee
Here one of the two couplings oscillates (Fourier components $g_{n,1}=(\delta_{n,1}+\delta_{n,-1})/2$) while the other is constant ($g_{n,2}=\delta_{n,0}$). Despite its simplicity, we also expect here the induced ratchet-like phenomenon,
which is one of the key ingredients {to obtain cooling by modulating the system-baths couplings}.
The spectral densities of the two baths are again Ohmic  with damping $\gamma_1$ and $\gamma_2$. Also in this case we define an effective damping strength $\gamma={\rm Max}[\gamma_1,\gamma_2]$ and the dimensionless parameter that governs the asymmetry  $\kappa= \gamma_1/\gamma_2$. 

The choice~\eqref{g_num} is also motivated by the fact that, for $\kappa\to 0$, the unconstrained model falls in the same universality class of the one in Eq.~\eqref{scelta_vincolo}. Indeed,  as shown in Appendix~\ref{loris_expansion}, up to linear order in $\kappa$ the heat currents  are equal  to  the perturbative expressions given  in Eqs.~\eqref{int1}-\eqref{int2}.  More generally, we have evaluated, at any order in $\kappa$, the heat currents, by first solving, via exact diagonalization, the algebraic equations~\eqref{floquet1} for the Floquet coefficients $G_m(\omega)$. Indeed, in this general case, several Floquet components, other than the static one, will give finite contributions. The heat currents are then obtained by inserting the results of $G_m(\omega)$ and the couplings~\eqref{g_num} into the general expression in Eq.~\eqref{jaaverage}. 

The numerical results are reported in Fig.~\ref{fig:5} for different values of $\kappa$ (red triangles in the plots). In Fig.~\ref{fig:5}(a) we show 
${\rm Max}[J_1]/P$, that is, 
the maximum positive value of $J_1$ normalized to the corresponding supplied power, in the isothermal case, in analogy with 
Fig.~\ref{fig:2}(b). Our results demonstrate that it is possible, also in this case, to obtain a positive value of $J_1$, and thus cooling induced by {suitably engineered} 
temporal modulation of the driven couplings. Here, the effect extends in a wider region of the asymmetry parameter $\kappa$ and still a finite (although small) effect is visible also for $\kappa \geq 1$. The dashed line in the plot represents the asymptotic value obtained in the perturbative regime at $\kappa \ll 1$, that corresponds to the value reported also in Fig.~\ref{fig:2}(b).
In Fig.~\ref{fig:5}(b) we consider finite thermal gradients $\Delta T\neq 0$, looking for refrigeration properties. Here, we report the maximum value of $\eta/\eta_C$ in percentage. Again, from this figure one can deduce that also refrigeration associated to the $\nu=1$ reservoir is a robust feature beyond the constraint~\eqref{vincolo}, and the optimal working regime (within this universality class) is obtained for small values of $\kappa$, i.e. in the case of strong asymmetry between the two bath couplings (but only in one direction, namely for $\gamma_1\ll\gamma_2$). This example proves the robustness of the discussed phenomenon beyond the particular choice of the time-dependent drives and interestingly it opens the possibility to study more complicated situations where the refrigeration response 
could be improved.
 
\subsection{Dynamical heat rectification}
\begin{figure}
\includegraphics[width=0.95\linewidth]{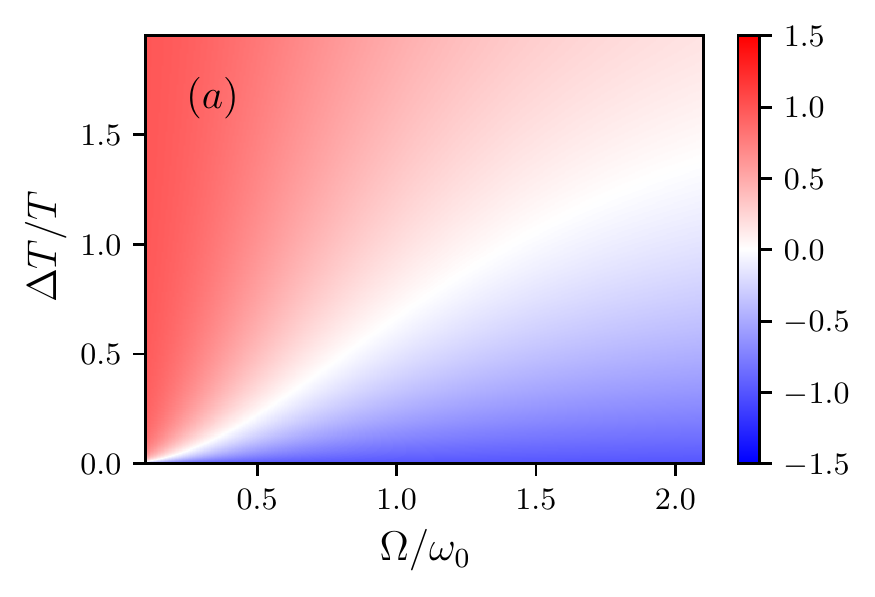}
\includegraphics[width=0.95\linewidth]{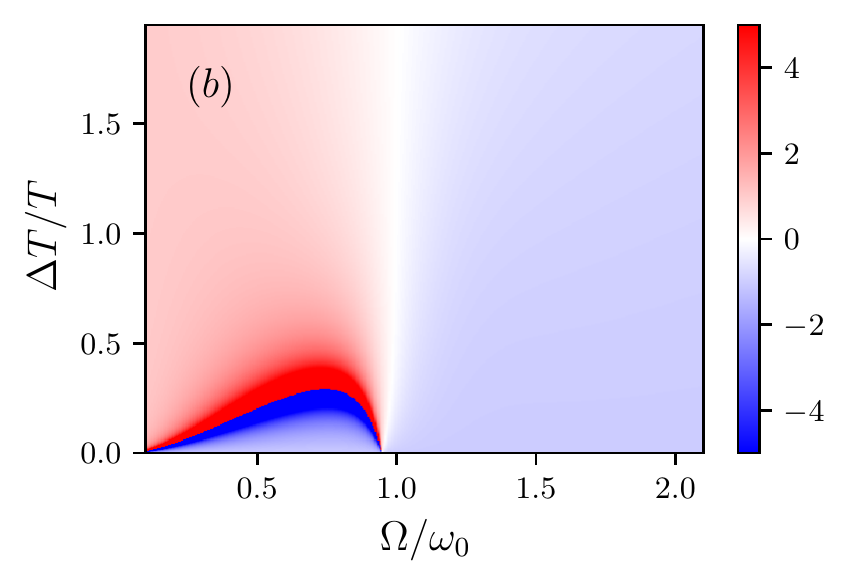}
\caption{\label{fig:6}
Dynamical heat rectification. Density plots of the ratio $R$ in Eq.~\eqref{def_r} between forward and backward heat currents as a function of external frequency $\Omega/\omega_0$ and normalized temperature gradient $\Delta T/T$. Panel (a): classic regime, with average temperature $T=10\omega_0$. Panel (b):  deep quantum regime with $T=0.15\omega_0$. Other parameters are $\kappa=0.01$ and $\gamma=0.01\omega_0$.}
\end{figure}
Before closing, we comment on another relevant aspect for quantum devices, i.e. the possibility to achieve rectification of heat current flows\cite{Giazotto2006, p14, Vannucci15, segal05, sanchez17, sanchez18, sanchez21, bours19, motz, bhandari21, flindt21}. It has been shown recently that heat rectification can be obtained in a linear system (as a QHO) by relying on external time-dependent forces~\cite{campeny19, peyrard}. Here we demonstrate that 
a dynamically induced heat rectification is also present without additional external fields but only in the presence of time dependent system-baths couplings. To this end, we focus on the heat current $J_1$, associated to the $\nu=1$ reservoir, in presence of a thermal gradient $\Delta T$ on top of an average temperature $T$. In order to quantify heat rectification, we define two 
configurations with interchanged temperatures, namely
\beq
\label{j_fb}
\!\!\!\!\!\!\!&&J_1^{\rm f}(T,\Delta T)= J_1( T_1=T+\Delta T/2, T_2=T-\Delta T/2),\nonumber\\
\!\!\!\!\!\!\!&&J_1^{\rm b}(T,\Delta T)=J_1(T_1=T - \Delta T/2,T_2=T+\Delta T/2).
\eeq
A useful figure of merit for rectification is then represented by the ratio
\be
\label{def_r}
R\equiv {-}\frac{J_1^{\rm f}(T,\Delta T)}{J_1^{\rm b}(T,\Delta T)}.
\ee
Here, we adopted the standard notation ${\rm f/b}=$ forward/backward, typically used in the presence of only a static thermal gradient\cite{Giazotto2006, p14, Vannucci15}, where
forward/backward represent  the direction of the heat fluxes{ and $R\ge 0$}. 
The value {$R=1$} indicates no heat rectification.
Notice that in our dynamically driven case, forward/backward does not necessarily imply a fixed direction of flow. Indeed, it is possible to have both heat currents  $J_1^{\rm f}$ and $J_1^{\rm b}$ flowing in the same direction with then
{$R<0$}. We have already met this situation in the isothermal case $\Delta T=0$, where a finite heat current {is in general present in spite of the  lack of thermal gradient, due to the asymmetric modulation of the couplings.
This is a kind of Peltier effect, but of dynamical origin.}
In this case forward and backward heat currents coincide with  
{$R=-1$}. In general,  by varying the temperature gradient and the driving frequency $\Omega$, the amplitudes and direction of the two heat currents change, resulting in 
{$|R|\neq 1$}. The situation where the forward (backward) configuration is completely blocked is indicated by $|R|\to 0$ or $|R|\to\infty$, respectively. 

The ratio $R$ is reported in Fig.~\ref{fig:6} as a function of the driving  frequency $\Omega/\omega_0$ and normalized temperature gradient $\Delta T/T$. As an example we  evaluated $R$  in the perturbative regime with $\kappa \ll 1$, using Eqs.~\eqref{int1}. Figure~\ref{fig:6}(a) shows the classical regime $T\gg\omega_0$. Here, for $\Delta T\to 0$, the dynamical Peltier contribution is the dominant  one: both $J_1^{\rm f}$ and $J_1^{\rm b}$ start negative  with
{ $R\to -1$}. Increasing $\Delta T$, 
$J_1^{\rm b}$ assumes  larger negative values, while  $J_1^{\rm f}$ decreases, until it changes sign. Therefore it is possible to block the heat current in the forward configuration ($R=0$, white regions in the density plot). Notice that the reverse situation of blocking the backward heat current ($|R|\to \infty$) is here never possible: indeed in Fig.~\ref{fig:6}(a) it is always $|R|\leq 1$. 

A much more versatile situation, instead, can be obtained in the quantum non Markovian regime at low average temperature $T\ll \omega_0$, as shown in Fig.~\ref{fig:6}(b). Here, we have two non overlapping regions which correspond to two orthogonal cases. The first one, with $\Omega\approx \omega_0$ (white area in the plot), has $R\to 0$,  signalling full blocking of $J_1^{\rm f}$, similar to the classical case. On the other hand,  in the second region (left bottom part of the density plot) one can achieve the full  block of the backward heat current $J_1^{\rm b}$ ($|R|\to\infty$).  Indeed, increasing  $\Delta T/T$ the backward current $J_1^{\rm b}$ changes sign passing from positive to negative values along the line situated between the two strong  blue and red color regions. Meanwhile $J_1^{\rm f}$ remains always positive. 
Importantly, these features are  present in the same parameter regions with $\Omega <\omega_0$ where finite refrigeration is obtained in response to the driven couplings (see Fig.~\ref{fig:4}(a)). 

In summary, {our system is much more versatile in the quantum regime},
since  in such a case it is possible,  by modulating the system-bath couplings, to create a heat rectifier which can switch by blocking either forward or backward current  by simply varying the driving frequency $\Omega$.

\section{Conclusions}
\label{conclusions}
{The extension of thermodynamics to small, quantum systems, challenges the usual paradigms of traditional thermodynamics,
like local thermal equilibrium, weak system-reservoir coupling, and Markovianity. 
As usual when facing the quantum world, even the most intuitive concepts should be carefully reexamined. 
For instance, one might reasonably argue that a purely dissipative effect is associated to the 
switching on/off the couplings to reservoirs in a nonadiabatic way, as required in any finite-time thermodynamic cycle.
In contrast, here we have shown that basic thermodynamic tasks can be performed by periodically modulating in a suitable way 
only the couplings to the baths. Indeed, asymmetric baths can be suitably engineered to induce cooling, refrigeration, 
and ideal heat rectification along a direction that can be reversed simply by tuning the modulation frequency. }

{We have described the quantum baths by the Caldeira-Leggett model, so that the system's dynamics and thermodynamics
can be investigated without resorting to the overdamped limit, to suitable master equations, or other approximations.
The usefulness of this general framework has been here tested for the  case where the system is a single harmonic oscillator,
the bath is Ohmic and only the system-baths couplings are time-dependent. On the other hand, our study paves the way to 
several possible generalizations. For instance, since the most intriguing results have been obtained in the non Markovian,
quantum regime, one could consider non Ohmic baths where non Markovian effects are present also in the classical,
high-temperature regime, in order to disentangle the relevance of non Markovian and quantum effects. 
Moreover, one could consider the joint effect of modulating the baths couplings and in addition driving the system, to investigate 
whether these external controls could cooperate in order to enhance the performance of refrigeration.
The same questions could be addressed for heat engines, and in both cases the developed formalism is ideally
suited to apply optimal control techniques~\cite{simone, mari, plastina, noe}. In particular, it would be interesting to reconsider
the results recently obtained~\cite{cangemi21} for isothermal heat engines, where in the antiadiabatic limit
the ideal efficiency is approached with finite output power and vanishingly small relative power fluctuations.
The intriguing question here is whether the simultaneous achievement of the three desiderata of a heat engine 
(efficiency close to the Carnot efficiency, high output power and constancy of the power output)
is possible also for standard heat engines operating with two or more heat baths at different temperatures. 
Further generalizations of our approach could be obtained by considering a more complex working medium\cite{Pekola21, jurgen1, jurgen2, gaspa, leitch, milne}, 
like coupled oscillators and, with a considerably higher numerical effort, nonlinear oscillators and 
qubit systems. 
}

\appendix

\section{Out of equilibrium Green function and Floquet coefficients}
\label{Floquet}
In this part we derive the  algebraic set of equations for the Floquet coefficients $\tilde{ G}_m(\omega)$ quoted in Eq.~(\ref{floquet1}).
We start from the differential equation in Eq.~(\ref{eomgreen}) written in the following compact form
\begin{eqnarray}
\label{appeomgreen}
&&\frac{\partial^2}{\partial t^2}{G}(t,t') + \omega_0^2 G(t,t')+\int_{-\infty}^{+\infty}\!\!\mathrm{d}s\Big[u(t,s)
G(s,t')\nonumber\\
&&+v(t,s)\frac{\partial}{\partial s}{G}(s,t')\Big] =\delta(t-t').
\end{eqnarray}
Here, we introduced the bath kernels
\beq
\label{kdefinition}
u(t,s)&=&\sum^N_{\nu=1}g_\nu(t)\gamma_\nu(t-s)\frac{d}{ds}g_\nu(s)\\
v(t,s)&=&\sum^N_{\nu=1}g_\nu(t)\gamma_\nu(t-s)g_\nu(s)
\eeq
expressed  in terms of the periodic couplings
\be
\label{appg}
g_{\nu}(t)=\sum_{n=-\infty}^{+\infty}g_{n,\nu} e^{-in\Omega t},\quad \Omega=\frac{2\pi}{{\cal T}}.
\ee

We remind that at long times it holds the property 

\be
\label{a1}
G(t,t')=\int_{-\infty}^{+\infty}\frac{\mathrm{d}\omega}{2\pi}e^{-i\omega(t-t')}\tilde{ G}(t,\omega),
\ee
with $\tilde{G}(t,\omega)$ a periodic function of $t$ with period ${\cal T}$:
\be
\label{a2}
\tilde{G}(t,\omega)=\sum_{m=-\infty}^{+\infty}\tilde{ G}_m(\omega) e^{-im\Omega t}.
\ee
A similar representation can be carried out also for  the kernels $u(t,s)$ and $v(t,s)$ 
\be
\label{apptransform2}
u(t,s)=\int_{-\infty}^{+\infty}\frac{\mathrm{d}\omega}{2\pi}e^{-i\omega(t-s)}\tilde{u}(t,\omega)
\ee
\be
\label{apptransform3}
v(t,s)=\int_{-\infty}^{+\infty}\frac{\mathrm{d}\omega}{2\pi}e^{-i\omega(t-s)}\tilde{v}(t,\omega)
\ee
with $\tilde{u}(t,\omega)$  and $\tilde{v}(t,\omega)$ given by
\be
\label{appseries1}
\tilde{u}(t,\omega)=\sum_{m=-\infty}^{+\infty}\tilde{u}_m(\omega) e^{-im\Omega t}.
\ee
and
\be
\label{appseries11}
\tilde{v}(t,\omega)=\sum_{m=-\infty}^{+\infty}\tilde{v}_m(\omega) e^{-im\Omega t}.
\ee
Now, we transform the differential equation (\ref{appeomgreen}) into a set of coupled algebraic equations. This can be done by inserting into Eq.~(\ref{appeomgreen}) the representations (\ref{a1}-\ref{a2}) and those for the kernels in Eqs.~(\ref{apptransform2}-\ref{appseries11}).
As a representative example we quote the expression  for the term associated to the damping contribution 
\be
{\cal D}(t,t')=\int_{-\infty}^{+\infty}\!\!\mathrm{d}s\Big[ u(t,s)
G(s,t')+v(t,s)\frac{\partial}{\partial s}{G}(s,t')\Big].
\ee
We have
\be
{\cal D}(t,\! t')\!\!=\!\!\int_{-\infty}^{+\infty}\!\!\!\frac{\mathrm{d}\omega}{2\pi}e^{-i\omega(t-t')}\!\!\!\!\sum_{m=-\infty}^{+\infty}\!\!\!\!
\tilde{ G}_{m}(\omega)\tilde{k}(t,\omega+m\Omega) e^{-im\Omega t}\nonumber\\
\ee
with
\be
\tilde{k}(t,\omega)=\tilde{u}(t,\omega)-i\omega\tilde{v}(t,\omega).\ee
Expressing $\tilde{k}(t,\omega)$ with the series (\ref{appseries1}-\ref{appseries11}) we obtain 
\beq
&&{\cal D}(t,t')=\int_{-\infty}^{+\infty}\frac{\mathrm{d}\omega}{2\pi}e^{-i\omega(t-t')}\sum_{m_1=-\infty}^{+\infty}
\sum_{m_2=-\infty}^{+\infty}\nonumber\\
&&\tilde{ G}_{m_1}(\omega)\tilde{k}_{m_2}(\omega+m_1\Omega) e^{-i(m_1+m_2)\Omega t},
\eeq
with 
\be
\tilde{k}_{m}(\omega)=\tilde{u}_{m}(\omega)-i\omega\tilde{v}_{m}(\omega).
\ee
Explicitly, we have
\be
\label{appkappan}
\tilde{k}_m(\omega)=-i\sum^N_{\nu=1}\sum_{n=-\infty}^{+\infty}g_{n,\nu}g_{m-n,\nu}\cdot(n\Omega+\omega)\tilde\gamma_\nu(\omega+n\Omega)
\ee
where
\be
\tilde\gamma_\nu(\omega)=\int_{-\infty}^{+\infty}\!\!\mathrm{d}t e^{i\omega t}\gamma_\nu(t).
\ee
 
Following similar steps for all the terms in Eq.~(\ref{appeomgreen}) we obtain the set of algebraic equations 
\beq
\label{a3}
&&[\omega_0^2-(\omega+m\Omega)^2]\tilde{ G}_m(\omega)+\nonumber\\
&&+\sum_{n=-\infty}^{+\infty}
\tilde{k}_n(\omega+(m-n)\Omega) \tilde{ G}_{m-n}(\omega)=\delta_{m,0}
\eeq

By introducing now  the ''static" retarded Green function in spectral domain
\be
\label{chi_app}
{\chi}_0(\omega)=-\frac{1}{\omega^2-\omega_0^2-\tilde{k}_0(\omega)},
\ee
which contains the $n=0$ component of the bath kernel $\tilde{k}_n(\omega)$ we rewrite Eq.~(\ref{a3}) in a compact form

\beq
&&\tilde{ G}_m(\omega)+\chi_0(\omega+m\Omega)\!\!\!\!\!\!\!\sum_{n=-\infty,n\neq 0}^{+\infty}\!\!\!\!\tilde{k}_n(\omega+(m-n)\Omega) \tilde{ G}_{m-n}(\omega)\nonumber\\
&&=\chi_0(\omega)\delta_{m,0}
\eeq
as reported in the main text.  

\section{Explicit expressions for average heat currents}
\label{heatpowerapp}
In this Appendix we derive the explicit expressions for the average heat current $J_\nu$ quoted in Eq.~(\ref{jaaverage}). 
We start by considering Eq.~(\ref{appja}) for the time dependent heat current $J_\nu(t)$ and we separate it into two contributions
\be 
J_\nu(t)=J_\nu^{(a)}(t)+J_\nu^{(b)}(t)
\ee
 where
\beq
\label{appjab}
&&J_{\nu}^{(a)}(t)= -\frac{g_\nu(t)}{m}\int_{-\infty}^{+\infty}\!\!\!\!\!\!\!\mathrm{d}t' G(t,t')g_{\nu}(t')\dot{\cal L}_\nu^{(+)}(t-t')\nonumber\\
&&J_{\nu}^{(b)}(t)= -\frac{2g_\nu(t)}{m^2}{\int_{-\infty}^{t}}\!\!\!\!\mathrm{d}s\int_{-\infty}^{+\infty}\!\!\!\!\!\!\mathrm{d}t_1\int_{-\infty}^{+\infty}\!\!\!\!\!\!\!\mathrm{d}t_2\, g_\nu(s)\dot{\cal L}_\nu^{(-)}(t-s)\nonumber\\
&&{\times} G(t,t_1)G(s,t_2)\sum_{\nu_1=1}^{N}g_{\nu_1}(t_1)g_{\nu_1}(t_2){\cal L}_{\nu_1}^{(+)}(t_1-t_2).
\eeq
We recall that the average heat currents $J_\nu$ are obtained from $J_{\nu}^{(a/b)}(t)$ after performing  the cycle average over the period ${\cal T}$
\be
\label{averageapp}
{{J}}_\nu^{(a/b)}=\frac{1}{{\cal T}}\int_{0}^{{\cal T}}\!\!\mathrm{d}t J_\nu^{(a/b)}(t),
\ee
with $J_\nu=J_\nu^{(a)}+J_\nu^{(b)}$ .

To proceed further, we rewrite $J_{\nu}^{(a/b)}(t)$ upon a change of variables as 
\beq
\label{appj1}
J_{\nu}^{(a)}(t)&=& -\int_{-\infty}^{+\infty}\!\!\!\!\!\mathrm{d}\tau\, \dot{\cal L}_\nu^{(+)}(\tau)M_\nu^{(a)}(t,t-\tau)\nonumber\\
J_{\nu}^{(b)}(t)&=&-\int_{0}^{+\infty}\!\!\!\!\!\mathrm{d}\tau\, \dot{\cal L}_\nu^{(-)}(\tau)M_\nu^{(b)}(t,t-\tau)
\eeq
with

\beq
\label{appjM}
&&M_\nu^{(a)}(t,t-\tau)=\frac{1}{m}g_\nu(t) g_{\nu}(t-\tau)G(t,t-\tau)\nonumber\\
&&M_\nu^{(b)}(t,t-\tau)=\frac{2}{m^2}g_\nu(t)g_\nu(t-\tau)\int_{-\infty}^{+\infty}\!\!\!\!\!\!\!\mathrm{d}t_1\!\!\int_{-\infty}^{+\infty}\!\!\!\!\!\!\!\mathrm{d}t_2\, 
G(t,t_1)\nonumber\\
&&{\times}G(t-\tau,t_2)\sum_{\nu_1=1}^{N}g_{\nu_1}(t_1)g_{\nu_1}(t_2){\cal L}_{\nu_1}^{(+)}(t_1-t_2).
\label{appjb1}
\eeq
Notice that the $t$ dependence  is now only present in the functions $M^{(a/b)}(t,t-\tau)$. Therefore  the average  (\ref{averageapp}) is 
\beq
\label{appjaverage}
{ J}_{\nu}^{(a)}&=& -\int_{-\infty}^{+\infty}\!\!\!\!\!\mathrm{d}\tau\, \dot{\cal L}_\nu^{(+)}(\tau)M_\nu^{(a)}(\tau)\nonumber\\
{ J}_{\nu}^{(b)}&=&-\int_{0}^{+\infty}\!\!\!\!\!\mathrm{d}\tau\, \dot{\cal L}_\nu^{(-)}(\tau)M_\nu^{(b)}(\tau)
\eeq
with
\be
\label{appMaverage}
{{M}}_\nu^{(a/b)}(\tau)=\frac{1}{{\cal T}}\int_{0}^{{\cal T}}\!\!\mathrm{d}t M_\nu^{(a/b)}(t,t-\tau).
\ee
We now evaluate ${{M}}_\nu^{(a/b)}(\tau)$. First we insert into Eq.~(\ref{appjM}) the representations (\ref{serieseg}) and (\ref{transform}) obtaining
\beq
&&M_\nu^{(a)}(t,t-\tau)=\!\!\!\!\!\!\sum_{m_1=-\infty}^{+\infty}\!\!\!\!e^{-im_1\Omega t}
\sum_{n_1,n_2=-\infty}^{+\infty}\!\!\!\!\!\!
g_{n_1,\nu}g_{n_2,\nu}\nonumber\\
&&{\times}\int_{-\infty}^{+\infty}\!\!\frac{\mathrm{d}\omega}{2\pi m}\tilde{ G}_{m_1-(n_1+n_2)}(\omega)e^{i(n_2 \Omega-\omega)\tau}\\\nonumber\\
&&M_\nu^{(b)}(t,t-\tau)\!=\!\!\!\!\!\!\!\sum_{m_1,m_2=-\infty}^{+\infty}\!\!\!\!\!\!\!e^{-im_1\Omega t}\!\!\!\!\!\!\!
\sum_{n_1\cdots n_4=-\infty}^{+\infty}\!\!\!\!\!\!g_{n_1,\nu}g_{n_2,\nu}\!\!\!
\sum_{\nu_1=1}^{N}\!g_{n_3,\nu_1}g_{n_4,\nu_1}\nonumber\\
&&{\times}\int_{-\infty}^{+\infty}\!\!\frac{\mathrm{d}\omega}{\pi m^2}
\tilde{ G}_{m_1-(m_2+ n_{\rm tot})}(\omega)\tilde{ G}_{m_2}((n_3+n_4)\Omega-\omega)\nonumber\\\nonumber\\
&&{\times}\widetilde{\cal L}_{\nu_1}^{(+)}(\omega- n_3\Omega)e^{i[(n_2+n_3+n_4)+m_2]\Omega\tau} e^{-i\omega\tau},
\eeq
where $n_{\rm tot}=n_1+n_2+n_3+n_4$. In the above expressions we introduced  the Fourier transform of the symmetric and antisymmetric part of the bath correlators ${\cal L}^{(\pm)}_\nu(t)$ in Eq.~\eqref{correlator}. They are  defined as
\be
\label{appcorrelator1}
\widetilde{\cal L}^{(\pm)}_\nu(\omega)=\int_{-\infty}^{\infty}\!\!\mathrm{d}t\,{\cal L}^{(\pm)}_\nu(t)e^{i\omega t},
\ee
and they have an explicit form in terms of the bath spectral densities ${\cal J}_\nu(\omega)$
\beq
\label{appLs}
\widetilde{\cal L}^{(+)}_\nu(\omega)&=&{\cal J}_\nu(\omega)\coth(\frac{\omega}{2T_\nu})\\
\widetilde{\cal L}^{(-)}_\nu(\omega)&=&i{\cal J}_\nu(\omega),
\label{appLa}
\eeq
with ${\cal J}_\nu(\omega)=- {\cal J}_\nu(-\omega)$. We now  perform the cycle average (\ref{appMaverage}) which yields
\beq
&&{{M}}_\nu^{(a)}(\tau)=\!\!\!\!\!\!\sum_{n_1,n_2=-\infty}^{+\infty}\!\!\!\!\!\!
g_{n_1,\nu}g_{n_2,\nu}\int_{-\infty}^{+\infty}\!\!\frac{\mathrm{d}\omega}{2\pi m}e^{i(n_2\Omega-\omega)\tau}\nonumber\\
&&\tilde{G}_{-(n_1+n_2)}(\omega)
\\\nonumber\\
&&{M}_\nu^{(b)}(\tau)\!=\!\!\!\!\!\!\sum_{m_1=-\infty}^{+\infty}\!\!\!
\sum_{\,\,\,\,n_1\cdots n_4=-\infty}^{+\infty}\!\!\!\!\!\!g_{n_1,\nu}g_{n_2,\nu}\!\!\!
\sum_{\nu_1=1}^{N}\!g_{n_3,\nu_1}g_{n_4,\nu_1}\nonumber\\
&&{\times}\int_{-\infty}^{+\infty}\!\!\frac{\mathrm{d}\omega}{\pi m^2}
\tilde{ G}_{-(m_1+n_{\rm tot})}(\omega)\tilde{ G}_{m_1}((n_3+n_4)\Omega-\omega)\nonumber\\\nonumber\\
&&{\times}\widetilde{\cal L}_{\nu_1}^{(+)}(\omega-n_3\Omega)e^{i[(n_2+n_3+n_4)+m_1]\Omega\tau} e^{-i\omega\tau}.
\eeq

Inserting these expressions into Eq.~(\ref{appjaverage}) we perform the $\tau$ integrals by using (\ref{appLs}), (\ref{appLa}). Notice that ${M}_\nu^{(b)}(\tau)={M}_\nu^{(b)}(-\tau)$. The final result for the average heat currents, once summed the two contributions, is reported in Eq.~\eqref{jaaverage} in the main text. 

Starting from Eq.~(\ref{Pexpression}) and following similar steps we obtain also the average power associated to the $\nu$-th bath:
\beq
\label{Paverage}
&&P_\nu=\Omega\!\!\!\!\sum_{n_1,n_2=-\infty}^{+\infty}\!\!\!\!\!\!
n_1g_{n_1,\nu}g_{n_2,\nu}\int_{-\infty}^{+\infty}\!\!\frac{\mathrm{d}\omega}{2\pi m}\Big\{i{\cal J}_\nu(\omega)\nonumber \\&&
{\times}  \coth(\frac{\omega}{2T_\nu}) \tilde{ G}_{-(n_1+n_2)}(\omega+n_2\Omega)
\\
&&\nonumber\\
&&+\sum_{\nu_1=1}^{N} \sum_{m_1=-\infty}^{+\infty}\sum_{n_3, n_4=-\infty}^{+\infty}\!g_{n_3,\nu_1}g_{n_4,\nu_1}\!\frac{{\cal J}_{\nu_1}(\omega)}{m}\coth(\frac{\omega}{2T_{\nu_1}})\nonumber\\&&\nonumber\\
&&{\times} {\cal J}_{\nu}(\omega-\Omega(n_2+n_4+m_1))\tilde{ G}_{m_1}(-\omega+n_4\Omega)\nonumber\\&&\nonumber\\
&&{\times} \tilde{ G}_{-(n_{\rm tot}+m_1)}(\omega+n_3\Omega)\Big\}.
\eeq

\section{Average heat currents in the classical regime}
\label{classic}
Here, we demonstrate that in the classical regime ($T\gg\omega_0$), the average heat currents for isothermal baths ($T_\nu=T$) are always dissipative ($J_\nu<0$).
To this end, we focus on the case where the time-dependent couplings are linked by the constraint~\eqref{vincolo}. We then start from the  heat current expressions given in Eq.~(\ref{jaaveragebis}).
In the classical  limit  we substitute  $\coth(\frac{\omega}{2T})\to 2T/\omega$, obtaining
\beq
\label{Jclassical}
&&{{J}}_{\nu}= T\gamma_\nu\int_{-\infty}^{+\infty}\!\!\frac{\mathrm{d}\omega}{{\pi}}\,\Big\{
\!\sum_{n=-\infty}^{+\infty}\!\!\!\!
|g_{n,\nu}|^2\omega\Im\chi_0(\omega+n\Omega)\nonumber\\&& -\!\!\!\!\!\!\sum_{n_1,n_3,n_4=-\infty}^{+\infty}\!\!\!\!\!\!\!\!\!g_{n_1,\nu}g_{-(n_1+n_3+n_4),\nu}\sum_{\nu_1=1}^{N}\gamma_{\nu_1}g_{n_3,\nu_1}g_{n_4,\nu_1}\nonumber\\
&&[\omega+\Omega(n_1+n_3)]^2
\chi_{0}(-\omega+n_4\Omega)\chi_0(\omega+n_3\Omega)\Big\}.
\eeq
Note that these expressions are well-behaved at large $\omega$ and therefore we have safely taken the $\omega_c\to\infty$ limit. We remind that $\Im\chi_0(\omega)$ has the following property
\be
\label{appchi}
\Im{\chi}_0(\omega)=\gamma\omega |\chi_0(\omega)|^2,
\ee
with 
\be
{\chi}_0(\omega)=\frac{-1}{\omega^2-\omega_0^2+i\omega\gamma}.
\ee
Expression (\ref{Jclassical}) can be rewritten by performing, in the second term, the change of variables $[\omega+\Omega(n_1+n_3)]\to \omega$ and $n_1+n_2+n_3\to n_3$
and by exploiting the relation
\be
\label{linkgn}
\sum_{\nu=1}^{N}\gamma_\nu \sum_{n=-\infty}^{+\infty}g_{n,\nu}g_{-n+m,\nu}=\gamma\delta_{m,0}
\ee
which derives from the link \eqref{vincolo}. We have the final result 
\be
\label{finalJcl}
{{J}}_{\nu}=-T\gamma_\nu\gamma\Omega^2\!\!\!\!\sum_{n=-\infty}^{+\infty}\!\!\!\!\!\!n^2|g_{n,\nu}|^2
\int_{-\infty}^{+\infty}\!\!\frac{\mathrm{d}\omega}{\pi}
\,\omega^2|\chi_0(\omega)|^2.
\ee
which is always negative. We can then conclude that  in the classical regime for isothermal Ohmic baths, it is impossibile to obtain  any cooling effect.

\section{Explicit expressions in the perturbative regime}
\label{complex}
In this Appendix, we derive the perturbative expressions for the average heat currents given in Eqs.~\eqref{int1}-\eqref{int2}.

We start by commenting on the iterative procedure used in perturbative schemes in order to solve the algebraic equations Eq.(\ref{floquet1}) for the Floquet coefficients.
The procedure starts, as a zero-th step, by choosing $\tilde{ G}_m(\omega)=\delta_{m,0}\chi_0(\omega)$, then 
by inserting it in (\ref{floquet1}), one finds the first corrections 
\be
\label{floquetfirst}
\tilde{ G}_{m\not =0}(\omega)=-\chi_0(\omega+m\Omega)
\tilde{k}_m(\omega)\chi_{0}(\omega).
\ee
The second iteration is obtained by dressing $\tilde{G}_0(\omega)$ with
\beq
\label{floquetsecond}
&&\tilde{ G}_{0}(\omega)=\chi_0(\omega) +\chi_0(\omega)
\sum_{n\not= 0}\tilde{k}_n(\omega-n\Omega)\nonumber\\
&&{\times}\chi_0(\omega-n\Omega)\tilde{k}_{-n}(\omega)
\chi_{0}(\omega),
\eeq
and so on and so forth. Looking at the formal structure of this expansion, one can identify the physical regimes where it is possible to safely stop the iteration by considering only the solutions given in Eqs.~(\ref{floquetfirst}) and (\ref{floquetsecond}): either when $\tilde{k}_{m\neq 0}(\omega)\ll \tilde{k}_0(\omega)$, namely a perturbation around the static term $ \tilde{k}_0(\omega)$, or conversely, at high driving frequencies $\Omega$. 

In the following  we will apply this scheme. We start by considering  the perturbative expansions of $g_\nu(t)$  given in Eq.~(\ref{gpert}) with Fourier transforms:
\beq
\label{gpertfourier}
\!\!\!\!\!\!\!\!\!g_{n,1}&=&\frac{1}{2}[\delta_{n,1}+\delta_{n,-1}]\nonumber\\
\!\!\!\!\!\!\!\!\!g_{n,2}&=&\sqrt{\frac{\gamma}{\gamma_2}}[(1-\frac{\kappa}{4})\delta_{n,0}-\frac{\kappa}{8}(\delta_{n,2}+\delta_{n,-2})\!+\!{\cal O}(\kappa^2).
\eeq
With these functions the kernels $\tilde{k}_n(\omega)$ in Eq.~\eqref{kappan1} become
\beq
\label{kappanappbis}
\tilde{k}_n(\omega)&=&-i\sum_{\nu=1}^2\gamma_{\nu} \sum_{m=-\infty}^{+\infty}g_{m,\nu}g_{n-m,\nu}(\omega+m\Omega)=\nonumber\\
&=&-i\gamma\omega\delta_{n,0}+{\cal O}(\kappa^2).
\eeq

We now apply the iterative solutions in Eqs.~(\ref{floquetfirst})-(\ref{floquetsecond}), obtaining the following $\kappa$ expansion of the  Floquet coefficients:
\be
\label{floquetpert}
\tilde{ G}_{m}(\omega)=\chi_0(\omega)\delta_{m,0}+{\cal O}(\kappa^2).
\ee
Notice that this expression fulfill the constraint of Eq. (\ref{GG0}) up to linear order in $\kappa$. For this reason we can directly use the general expressions~(\ref{jaaveragebis}) for the heat currents, valid in the constrained case, 
by inserting Eqs.~(\ref{gpertfourier}) and (\ref{floquetpert}). 

We start with $J_1$. We observe the presence of the factor $\gamma_1=\kappa\gamma$ in front of the integral, this implies to evaluate all other terms
 at zero-th order in $\kappa$, namely putting $\nu_1=2$ and 
$g_{n,2}=\sqrt{\frac{\gamma}{\gamma_2}}\delta_{n,0}$. The result is 
\beq
\label{int1app}
&&J_{1}=\kappa\gamma \ \int_{-\infty}^{+\infty}\!\!\frac{\mathrm{d}\omega}{4\pi}\,\Big\{
-(\omega^2+\Omega^2)\Im\chi_{0}(\omega)\coth(\frac{\omega}{2T_{2}})\nonumber\\
&&+\omega^2 \Im\chi_0(\omega+\Omega)\coth(\frac{\omega}{2T_1})\Big\}+{\cal O}(\kappa^2), 
\eeq
as quoted in Eq.~\eqref{int1}.  

More cumbersome is the evaluation of $J_2$. Here, the first term in Eq.~(\ref{jaaveragebis}), called $J_2^{(a)}$,  is 
\be
\label{int2aapp}
\!J_{2}^{(a)}=\gamma(1-\frac{\kappa}{2})\!\!\!\int_{-\infty}^{+\infty}\!\!\frac{\mathrm{d}\omega}{2\pi}
\omega^2\Im\chi_{0}(\omega)\coth(\frac{\omega}{2T_{2}})\!+{\cal O}(\kappa^2)
\ee
and it contains also a zero-th order term, that, as we will see shortly, will be cancelled out from the remaining part of $J_2$. 
This one, called $J_2^{(b)}$, receives contributions  coming from both reservoirs ${\nu_1=1,2}$. For $\nu_1=1$ we again have  to put $g_{n,2}=\sqrt{\frac{\gamma}{\gamma_2}}\delta_{n,0}$ obtaining
\beq
\label{int2b1app}
\!\!\!\!\!&&J_{2}^{(b)}(\nu_1=1)=-\gamma\frac{\kappa}{2}\int_{-\infty}^{+\infty}\!\!\frac{\mathrm{d}\omega}{2\pi}\omega(\omega+\Omega)\coth(\frac{\omega}{2T_{1}})
\nonumber\\
&&{\times}\Im\chi_0(\omega+\Omega)\!+{\cal O}(\kappa^2).
\eeq
For $\nu_1=2$ the contribution $J_{2}^{(b)}(\nu_1=2)$ has also a zero term (opposite to the one of $J_{2}^{(a)}$) in addition to the linear one:
 \beq
\label{int2b2app}
\!\!\!\!\!&&J_{2}^{(b)}(\nu_1=2)=-\gamma(1-\kappa)\int_{-\infty}^{+\infty}\!\!\frac{\mathrm{d}\omega}{2\pi}\omega^2\coth(\frac{\omega}{2T_{2}})\nonumber\\
&&{\times}\Im\chi_0(\omega)\!+{\cal O}(\kappa^2).
\eeq
Summing up all these terms we arrive to $J_2=J_{2}^{(a)}+J_{2}^{(b)}(\nu_1=1)+J_{2}^{(b)}(\nu_1=2)$ with
\beq
\label{int2app}
\!\!\!&&{{J}}_{2}=\kappa\gamma \int_{-\infty}^{+\infty}\!\!\frac{\mathrm{d}\omega}{4\pi}\,\Big\{
\omega^2\Im\chi_{0}(\omega)\coth(\frac{\omega}{2T_{2}})\nonumber\\
\!\!\!&&-\omega(\omega+\Omega)\Im\chi_0(\omega+\Omega) \coth(\frac{\omega}{2T_1})\Big\}\!+{\cal O}(\kappa^2)
\eeq
which is the result quoted in Eq.~(\ref{int2}).\\

We now present the explicit evaluation of the expressions (\ref{int1app})-(\ref{int2app}).

First of all, we observe that $\chi_0(\omega)$ in Eq.~(\ref{chi_app}) can be decomposed as 
\be
\chi_0(\omega)=-\frac{1}{2\xi}\left[ \frac{1}{\omega+i\lambda_1}- \frac{1}{\omega+i\lambda_2}\right]
\ee
where
\beq
&& \lambda_1=\frac{\gamma}{2}+i\xi;\quad  \lambda_2=\frac{\gamma}{2}-i\xi\nonumber\\
&&\xi=\sqrt{\omega_0^2-\gamma^2/4}.
\eeq
Notice that for $\gamma <2\omega_0$ the roots are complex conjugate (underdamped oscillator), otherwise for  $\gamma >2\omega_0$ they are real (overdamped oscillator).
Hereafter, we will consider the first case, which is the most interesting in the situation under investigation.
In this regime, we have

\be
\label{chidecomposition}
\Im\chi_0(\omega)=\frac{\gamma}{4\xi}\left[\frac{1}{(\omega+i\lambda_1)(\omega-i\lambda_2)}-\frac{1}{(\omega+i\lambda_2)(\omega-i\lambda_1)}\right]
\ee
Notice that in the integrals for the average heat currents \eqref{int1}-\eqref{int2}, it is always present the function $\coth(\frac{\omega}{2T_\nu})$ which we now express as a series in the matsubara frequencies 
$\omega_{n,\nu}=2\pi n T_\nu$
\be
\label{cothseries}
\coth(\frac{\omega}{2T_\nu})=2T_\nu\Big[\frac{1}{\omega}+2\omega\sum_{n=1}^{+\infty}\frac{1}{\omega^2+\omega_{n,\nu}^2}\Big].
\ee
The integration will be performed in the complex plane using Cauchy method and considering a closed contour in the upper half plane. The poles are of two kinds: those given by the  
$\Im\chi_0(\omega)$ and $\Im\chi_0(\omega+\Omega)$, situated in $\omega=i\lambda_2, i\lambda_1$ and in $\omega= i\lambda_2-\Omega, i\lambda_1-\Omega$; and those given by the $\coth$-function, located at $\omega=i\omega_{n,\nu}$. Notice that there is no pole in $\omega=0$. Moreover, since in general we are dealing with two different temperatures $T_1$ and $T_2$ the associated poles will be placed in different positions and one should properly take care of this fact while considering the limit procedure $\rho \to \infty$ of the radius of the closed path.
We can then always split the result of the integral (\ref{int1}) into a sum of a contribution due the poles of the $\Im\chi_0$ and one due to the  poles of the $\coth$-function. 
Below, we explicitly evaluate $J_1$ and the average total power $P=-(J_1+J_2)$ from which one can infer also $J_2$. We can write
\beq
&&{{J}}_{1}=J_{1,\chi}+J_{1,\coth}\nonumber\\
&&P=P_{\chi}+P_{\coth}.
\eeq
The contributions of the poles of $\Im\chi_0$ are
\beq
\label{risultato}
&&J_{1,\chi}=\frac{\gamma\kappa}{8\xi}\Big\{(\Omega^2-\lambda_1^2)\coth(\frac{i\lambda_1}{2T_2})
-(\Omega^2-\lambda_2^2)\coth(\frac{i\lambda_2}{2T_2})\nonumber\\
&&+(i\lambda_2-\Omega)^2\coth(\frac{i\lambda_2-\Omega}{2T_1})-(i\lambda_1-\Omega)^2\coth(\frac{i\lambda_1-\Omega}{2T_1})\Big\}\nonumber\\
\eeq
and 
\beq
&&P_{\chi}=-\frac{\gamma\kappa\Omega}{8\xi}\Big\{\Omega\coth(\frac{i\lambda_1}{2T_2})
-\Omega\coth(\frac{i\lambda_2}{2T_2})\nonumber\\
&&-(i\lambda_2-\Omega)\coth(\frac{i\lambda_2-\Omega}{2T_1})+(i\lambda_1-\Omega)\coth(\frac{i\lambda_1-\Omega}{2T_1})\Big\}\nonumber\\
\eeq
The poles of the $\coth$-function instead give the following contributions
\beq
&&J_{1,\coth}=-i\gamma\kappa\Big\{ \sum_{n=1}^{+\infty}T_2(\Omega^2-\omega_{n,2}^2)\Im\chi_0(i\omega_{n,2})\nonumber\\
&&+\sum_{n=1}^{+\infty} T_1\omega_{n,1}^2\Im\chi_0(i\omega_{n,1}+\Omega)\Big\}
\eeq
and 
\beq
&&P_{\coth}=i\gamma\kappa\Omega\Big\{ \sum_{n=1}^{+\infty}T_2\Omega\Im\chi_0(i\omega_{n,2})\nonumber\\
&&+\sum_{n=1}^{+\infty} iT_1\omega_{n,1}\Im\chi_0(i\omega_{n,1}+\Omega)\Big\}
\eeq

We are then left to resum the above series. This can  be done thanks to the relative simple form of  $\Im\chi(\omega)$.
The typical series we need is:
\be
L_{a,b}=\sum_{n=1}^{+\infty}\frac{1}{(n+a)(n+b)}=\frac{1}{b-a}[\psi(b)-\psi(a)]-\frac{1}{ab}
\ee
with $\psi(x)$ the digamma function. After a long but standard procedure we arrive to the following exact results
\beq
&& J_{1,\coth} =i\frac{\gamma^2 \kappa}{4\xi}\Big\{\frac{\Omega^2}{4\pi^2 T_2}[L_{a,b}-L_{c,d}] - T_2 a^2 L_{a,b} + T_2 c^2 L_{c,d} \nonumber \\&& + T_1 \bar{a}^2 L_{\bar{a},\bar{b}} - T_1 \bar{c}^2 L_{\bar{c},\bar{d}} + \frac{\Omega \xi}{\pi^2 T_1} L_{\bar{b},\bar{d}}
\nonumber \\
&&+ \frac{i \xi}{\pi}\big[(\bar{b}-b)L_{b,\bar{b}} + (\bar{d}-d)L_{d,\bar{d}}+2\log(T_1/T_2)\big]\Big\}
\eeq

\noindent and 
\beq
&& P_{\rm coth} =-\frac{\gamma^2 \kappa}{4\xi}\Big\{\frac{i\Omega^2}{4\pi^2 T_2}[L_{a,b}-L_{c,d}]+\nonumber\\
&&-\frac{\Omega}{2\pi}\big[(\bar{d}-\bar{b})L_{\bar{b},\bar{d}} -\bar{a}L_{\bar{a},\bar{b}}+\bar{c}L_{\bar{c},\bar{d}}
\big]\Big\}
\eeq
where for notational convenience we have defined the quantities:
\beq
&&a=\frac{\lambda_1}{2\pi T_2};\quad \bar a=\frac{\lambda_1-i\Omega}{2\pi T_1}\nonumber\\
&&b=\frac{-\lambda_2}{2\pi T_2};\quad \bar b=\frac{-\lambda_2-i\Omega}{2\pi T_1}\nonumber\\
&&c=\frac{\lambda_2}{2\pi T_2};\quad \bar c=\frac{\lambda_2-i\Omega}{2\pi T_1}\nonumber\\
&&d=\frac{-\lambda_1}{2\pi T_2};\quad \bar d=\frac{-\lambda_1-i\Omega}{2\pi T_1}
\eeq

\section{Perturbative regime for the unconstrained case}
\label{loris_expansion}

In this Appendix, we demonstrate that  the unconstrained  coupling fields considered  in Eq.~(\ref{g_num}) belong to the same perturbative class as the constraint one.

Let us start by recalling  the  Fourier transforms of the coupling fields:
\beq
\label{gpertfourierL}
\!\!\!\!\!\!\!\!\!g_{n,1}&=&\frac{1}{2}[\delta_{n,1}+\delta_{n,-1}]\nonumber\\
\!\!\!\!\!\!\!\!\!g_{n,2}&=&\delta_{n,0}.
\eeq
With these functions the kernels $\tilde{k}_n(\omega)$ in Eq.~\eqref{kappan1} are exactly given by
\be
\label{kappanappbisL}
\!\tilde{k}_n(\omega)\!=\!-i\gamma_2\Big[\omega(1+\frac{\kappa}{2})\delta_{n,0}+\frac{\kappa}{4}[(\omega+\Omega)\delta_{n,2}+(\omega-\Omega)\delta_{n,-2}]\Big],
\ee
where we remind the definition of the effective asymmetry $\kappa=\gamma_1/\gamma_2$ and effective damping $\gamma={\rm Max}[\gamma_1,\gamma_2]$. In the following we consider the perturbative regime $\kappa\to 0$.
Inserting (\ref{kappanappbisL}) and (\ref{gpertfourierL}) into the algebraic equations~(\ref{floquet1})
the corresponding  Floquet coefficients are:
\beq
\label{floquetpertL}
&&\tilde{ G}_{0}(\omega)=\chi_0(\omega)[1+i\frac{\omega\gamma\kappa}{2}\chi_0(\omega)]+{\cal O}(\kappa^2)\nonumber\\
&&\tilde{ G}_{\pm2}(\omega)=i\frac{\gamma\kappa}{4}(\omega\pm \Omega)\chi_0(\omega\pm 2\Omega)\chi_0(\omega)+{\cal O}(\kappa^2)\nonumber\\
&&\tilde{ G}_{|m|>2}(\omega)={\cal O}(\kappa^2)
\eeq
with ${\chi}_0(\omega)=-1/[\omega^2-\omega_0^2+i\gamma\omega]$ the bare susceptivity. Notice that for symmetry reasons we always  have $\tilde{ G}_{2m+1}(\omega)=0$.
We now evaluate the general expressions~(\ref{jaaverage}) for the heat currents taking into account Eqs.~(\ref{gpertfourierL}) and (\ref{floquetpertL}). We start with  $J_1$, which already  contains the factor $\gamma_1=\kappa\gamma$ in front of the integral. This implies to evaluate all other terms at zero order, namely putting in the first term of the integral $\tilde{ G}_{n_1+n_2}(\omega+n_2\Omega)=\chi_0(\omega+n_2\Omega)\delta_{n_1,-n_2}$ and in the second part $\nu_1=2$, and 
$\tilde{ G}_{m_1}(-\omega+n_4\Omega)\tilde{ G}_{-(n_{\rm tot}+m_1)}(\omega+n_3\Omega)=\chi_0(-\omega+n_4\Omega)\chi_0(\omega+n_3\Omega)\delta_{m_1,0}\delta_{n_{\rm tot},0}$. With this procedure we obtain exactly the perturbative results of the constraint case given in Eq.~\eqref{int1app}.
We now consider $J_2$.  The first term in Eq.~(\ref{jaaverage}), called $J_2^{(a)}$,  has $n_1=n_2=0$ and can be written as
\be
\label{int2aappL}
J_{2}^{(a)}\!=\!-i\gamma\!\!\int_{-\infty}^{+\infty}\!\!\frac{\mathrm{d}\omega}{2\pi}
\omega^2\coth(\frac{\omega}{2T_{2}})\tilde{ G}_{0}(\omega).
\ee
Replacing into $\tilde{ G}_{0}(\omega)$ the low $\kappa$ expansion (\ref{floquetpertL}) we have
\beq
\label{int2aappLbis}
&&J_{2}^{(a)}=\gamma\!\int_{-\infty}^{+\infty}\!\!\frac{\mathrm{d}\omega}{2\pi}
\omega^2\coth(\frac{\omega}{2T_{2}})\Big\{\Im\chi_{0}(\omega)+\nonumber\\
&&+\frac{\omega\gamma\kappa}{2}[\Re^2\chi_{0}(\omega)-\Im^2\chi_{0}(\omega)]\Big\}+{\cal O}(\kappa^2).
\eeq
The second term, called $J_2^{(b)}$, has two contributions  coming from both reservoirs ${\nu_1=1,2}$. For $\nu_1=1$ we again need to evaluate all terms at zero order in $\kappa$ because  ${\cal J}_{\nu_1=1}(\omega)$ contains already $\gamma_1$. This implies $m_1=0$, with $\tilde{ G}_{0}(\omega)=\chi_0(\omega)$ and $n_3=-n_4$. 
The final expression for $J_{2}^{(b)}(\nu_1=1)$ is then equal to the one quoted in Eq.~(\ref{int2b1app}).
We are left to consider the last term with $\nu_1=2$ called $J_{2}^{(b)}(\nu_1=2)$.  Here, we have $n_1=n_2=n_3=n_4=0$ and up to linaer order in $\kappa$ also $m_1=0$.
The result is
\be
\label{int2b2appL2}
\!\!J_{2}^{(b)}(\nu_1=2)\!=\!-\gamma^2\!\!\int_{-\infty}^{+\infty}\!\!\frac{\mathrm{d}\omega}{2\pi}\omega^3\!\!\coth(\frac{\omega}{2T_{2}})|\tilde{ G}_{0}(\omega)|^2
\!+{\cal O}(\kappa^2).
\ee
By replacing 
\be
\gamma\omega|\tilde{ G}_{0}(\omega)|^2=\Im\chi_{0}(\omega)\Big[1-\omega\gamma\kappa\Im\chi_{0}(\omega)\Big]+{\cal O}(\kappa^2)
\ee
we obtain
\beq
\label{int2b2appL3}
\!\!\!\!\!&&J_{2}^{(b)}(\nu_1=2)=-\gamma\!\int_{-\infty}^{+\infty}\!\!\frac{\mathrm{d}\omega}{2\pi}
\omega^2\coth(\frac{\omega}{2T_{2}})\Im\chi_{0}(\omega)\nonumber\\
&&{\times}\Big\{1-\gamma\kappa\omega\Im\chi_{0}(\omega)\Big\}+{\cal O}(\kappa^2).
\eeq

Summing up all these terms we have  $J_2=J_{2}^{(a)}+J_{2}^{(b)}(\nu_1=1)+J_{2}^{(b)}(\nu_1=2)$ which is equal to the perturbative result (\ref{int2app}) obtained in the constrained case.


\begin{thebibliography}{99}
\bibitem{footnote:Heisenberg}
We denote instead as $O(t)$ an observable $O$ at time $t$ in the Heisenberg picture.
\end{thebibliography}
\end{document}